\documentclass[11pt,a4paper]{article}
\RequirePackage[l2tabu,orthodox,abort]{nag}

\usepackage{fixltx2e}

\usepackage[font=small]{caption}
\usepackage[text={159mm,245mm},headheight={14.5pt},centering]{geometry}

\usepackage{lmodern}
\usepackage{graphicx}
\usepackage{subfig}
\usepackage{booktabs}
\usepackage{multirow}
\usepackage{dsfont}
\usepackage[numbers,merge]{natbib}
\usepackage{simplewick}

\usepackage{tikz}
\usetikzlibrary{plotmarks,calc,decorations.markings,decorations.pathmorphing,decorations.pathreplacing,patterns,matrix}

\usepackage[intlimits]{amsmath}
\usepackage{amssymb}
\usepackage{mathtools}
\usepackage{slashed}
\usepackage{braket}

\usepackage{hyperref}

\usepackage{xspace}

\usepackage{accents}

\numberwithin{equation}{section}

\makeatletter
 \let\old@startsection=\@startsection
 \let\oldl@section=\l@section
 \renewcommand{\@startsection}[6]{\old@startsection{#1}{#2}{#3}{#4}{#5}{#6\mathversion{bold}}}
 \renewcommand{\l@section}[2]{\oldl@section{\mathversion{bold}#1}{#2}}
\makeatother

\renewcommand{\geq}{\geqslant}

\DeclareMathOperator{\str}{str}
\DeclareMathOperator{\Sym}{Sym}
\DeclareMathOperator{\tr}{tr}

\newcommand{\AdS}{\textup{AdS}}
\newcommand{\CFT}{\textup{CFT}}
\newcommand{\Sphere}{\textup{S}}
\newcommand{\Torus}{\textup{T}}
\newcommand{\CP}{\textup{CP}}
\newcommand{\Kthree}{\textup{K3}}

\newcommand{\Smat}{\mathcal{S}}
\newcommand{\Rmat}{\mathcal{R}}

\newcommand{\comm}[2]{[#1,#2]}
\newcommand{\acomm}[2]{\{#1,#2\}}

\newcommand{\alg}[1]{\mathrm{#1}}
\newcommand{\grp}[1]{\mathrm{#1}}

\newcommand{\algSU}{\alg{su}}

\newcommand{\algPSU}{\alg{psu}}
\newcommand{\grpPSU}{\grp{PSU}}

\newcommand{\gen}[1]{\mathbf{#1}}

\newcommand{\rep}[1]{\mathbf{#1}}

\newcommand{\Integers}{\mathbf{Z}}

\newcommand{\order}{\mathcal{O}}

\newcommand{\superN}{\mathcal{N}}

\newcommand{\ie}{\textit{i.e.}\xspace}
\newcommand{\eg}{\textit{e.g.}\xspace}

\newcommand{\ce}{\text{c.e.}}

\DeclareMathOperator{\sech}{sech}

\newcommand{\sL}{{\scriptscriptstyle\mathrm{L}}}
\newcommand{\sR}{{\scriptscriptstyle\mathrm{R}}}
\newcommand{\sP}{{\scriptscriptstyle+}}
\newcommand{\sM}{{\scriptscriptstyle-}}
\newcommand{\sPM}{{\scriptscriptstyle\pm}}

\newcommand{\sS}{\mbox{\tiny S}}

\newcommand{\aba}[1]{\mathcal{#1}}

\newcommand{\hslashslash}{%
  \raisebox{.9ex}{%
    \scalebox{.7}{%
      \rotatebox[origin=c]{17}{$-$}%
    }%
  }%
}

\renewcommand{\k}{
  {
   \vphantom{k}
   \ooalign{\kern-.05em\smash{\hslashslash}\hidewidth\cr$k$\cr}
   \kern-.025em
  }
}

\begin{document}

\thispagestyle{empty}
\begin{flushright}\footnotesize\ttfamily
DMUS-MP-21/05
\\
NORDITA 2021-023
\end{flushright}
\vspace{2em}

\begin{center}

{\Large\bfseries \vspace{0.2cm}
{\color{black} Protected states in $\AdS_3$ backgrounds from integrability}} 
\vspace{1.5cm}

\textrm{\large 
Suvajit Majumder${}^{*}$, \ 
Olof Ohlsson Sax${}^{\sharp}$, \ 
Bogdan Stefa\'nski, jr.${}^{*}$, \  Alessandro Torrielli${}^{\dagger}$
}

\vspace{2em}

\vspace{1em}
\begingroup\itshape
${}^{*}$ Centre for Mathematical Science, City, University of London
\\
Northampton Square, EC1V 0HB London, UK
\vspace{0.2cm}
\\
${}^{\sharp}$ Nordita, Stockholm University and KTH Royal Institute of Technology,
\\
Roslagstullsbacken 23, SE-106 91 Stockholm, Sweden
\vspace{0.2cm}
\\
${}^{\dagger}$ Department of Mathematics, University of Surrey
\\
 Guildford, GU2 7XH, UK
\par\endgroup

\vspace{2em}

\begingroup\ttfamily
suvajit.majumder@city.ac.uk,
olof.ohlsson.sax@gmail.com,
Bogdan.Stefanski.1@city.ac.uk,
a.torrielli@surrey.ac.uk
\par\endgroup

\end{center}

\vspace{2em}

\begin{abstract}\noindent 
We write down the Algebraic Bethe Ansatz for string theory on 
$\AdS_3\times\Sphere^3\times\Torus^4$ and $\AdS_3\times\Sphere^3\times\Kthree$ in its orbifold limits. We use it to determine the wave-functions of protected closed strings in these backgrounds and prove that their energies are protected to all orders in $\alpha'$. We further apply the ABA to find the wave functions of protected states of $\AdS_3\times\Sphere^3\times\Sphere^3\times \Sphere^1$ and its $\Integers_2$ orbifold.
Our findings match with protected spectrum calculations from supergravity, $\Sym^N$ orbifolds and apply to the complete moduli space of these theories, excluding orbifold blow-up modes for which further analysis is necessary.

\end{abstract}

\newpage

\overfullrule=0pt
\parskip=2pt
\parindent=12pt
\headheight=0.0in \headsep=0.0in \topmargin=0.0in \oddsidemargin=0in

\vspace{-3cm}
\thispagestyle{empty}
\vspace{-1cm}

\tableofcontents

\setcounter{footnote}{0}

\newpage

\section{Introduction}
Integrability is a powerful method for understanding certain holographic gauge/gravity models, because it allows for \textit{exact} computations in the `t Hooft coupling $\lambda$, or equivalently in $\alpha'$ (for a review see~\cite{Beisert:2010jr}). Key non-protected quantities, such as the dimension of  Konishi or twist-two operators~\cite{Gromov:2009bc,Bombardelli:2009ns,Gromov:2009tv,Arutyunov:2009ur,Gromov:2015wca}, can be computed to a very high level of precision, not just far higher than one could hope to do with direct Feynman graph techniques, but also interpolating between weak, intermediate and strong coupling. 

At the same time, integrability offers a novel way to prove non-renormalization theorems, since one can show exactly in $\lambda$ that certain BPS operators' dimensions do not receive quantum corrections. In the planar limit, integrable $\AdS_5\times\Sphere^5$ and $\AdS_4\times \CP^3$ backgrounds have only one protected multiplet for each value of global charges\footnote{See the review~\cite{Bajnok:2010ke} and references in it.}. Geometrically, this corresponds to the supersymmetric light-ray, or Berenstein-Maldacena-Nastase (BMN) geodesic~\cite{Berenstein:2002jq}, while in the gauge theory language protected operators are the (unique) ferromagnetic groundstates of the Minahan-Zarembo spin-chains~\cite{Minahan:2002ve}.\footnote{In the case of planar $\superN=4$ this is the $\tr(Z^J)$ family, while in ABJM theory it is $\tr((AB)^J)$.} 

The protected-spectrum of integrable $\AdS_3/\CFT_2$ backgrounds~\cite{Babichenko:2009dk,OhlssonSax:2011ms,Cagnazzo:2012se} is much richer, with several  multiplets for a given set of charges.  In the integrable formulation, these extra multiplets appear because  the worldsheet theory has fermionic massless excitations~\cite{Sax:2012jv}. Such excitations have zero energy in the zero-momentum limit but are not descendents of the BMN vacuum. Further, for each set of charges one can find solutions of the exact Bethe equations~\cite{Borsato:2016kbm,Borsato:2016xns} with the correct multiplicities to match the supergravity and $\Sym^N$ orbifold calculations~\cite{deBoer:1998kjm,Eberhardt:2017fsi}. Since  integrability methods are exact in $\alpha'$, they give $\AdS_3\times \Sphere^3\times \mathcal{M}_4$ non-renormalization theorems for \textit{all} half- and quarter-BPS states of 
$\mathcal{M}_4=\Torus^4$ and $\mathcal{M}_4=\Sphere^3\times\Sphere^1$, respectively~\cite{Borsato:2016kbm,Baggio:2017kza}. These backgrounds have a 20, respectively 2, dimensional moduli space on which integrability continues to hold~\cite{OhlssonSax:2018hgc}. From this we can immediately conclude that  half-, respectively quarter-BPS, states are protected across the whole moduli space, matching in particular the WZW point results~\cite{Dabholkar:2007ey,Eberhardt:2017fsi}. These findings are also in agreement with the non-renormalization results~\cite{Baggio:2012rr}  applicable to the $\mathcal{M}_4=\Torus^4$ case.

As in other integrable models, Bethe equations (BEs) are valid in the large worldsheet radius limit with generic states receiving wrapping corrections. These are especially important when the theory has massless modes~\cite{Abbott:2015pps,Abbott:2020jaa}. It is known that protected states do not receive wrapping corrections at leading-order~\cite{Baggio:2017kza} and it is likely that the argument can be generalised to all orders of wrapping using the exact massless TBA~\cite{Bombardelli:2018jkj,Fontanella:2019ury}. 

In this paper we find the Bethe eigenvectors for massless low-magnon excitations using the Algebraic Bethe Ansatz (ABA). Low-lying states in integrable field theories are analogous to conventional Minkowski spacetime string states with a few oscillator excitations above the vacuum, for example
\begin{equation}
\ket{v}\equiv\left(\alpha^i_{p_1}\right)^\dagger\left(S^{\dot{a}}_{p_2}\right)^\dagger\left(\tilde{\alpha}^{j}_{p_3}\right)^\dagger\left|0\right>.
\label{eq:flat-space-state}
\end{equation}
Each of the three magnons above sits in a short representation of the supersymmetry algebra $\alg{psu}(1|1)^4_{\ce}$ unbroken by the BMN vacuum.\footnote{The global symmetry of $\AdS_3\times\Sphere^3\times\Torus^4$ is $\alg{psu}(1,1|2)^2$  and the vacuum-preserving symmetry sub-algebra is  $\alg{p}(\alg{su}(1|1)^2)^2$. Since individual magnons do not satisfy the level-matching condition, they transform in representations of a triple central extension of $\alg{p}(\alg{su}(1|1)^2)^2$, which we denote by $\alg{psu}(1|1)^4_{\ce}$.} The indices $i$ and $\dot{a}$ are the usual transverse $\rep{8}_v$ and $\rep{8}_c$ indices of the light-cone $\alg{so}(8)$ algebra~\cite{Green:1987sp}, which now break-up into several representations of the (abelian) bosonic subalgebra of $\alg{psu}(1|1)^4_{\ce}$, depending on the mass $m$ of the magnon. The energy of the above state is the sum of the energies of the individual magnons
\begin{equation}
E(v)=E(p_1)+E(p_2)+E(p_3),
\end{equation}
 each of which satisfies the exact dispersion relation dictated by the shortening condition~\cite{Hoare:2013lja,Lloyd:2014bsa}
\begin{equation}\label{eq:disp-rel}
E(p_i)=\sqrt{(m_i+\k p_i)^2 +4 h^2 \sin^2\tfrac{p_i}{2}}.
\end{equation}
Above, $2\pi \k\in\Integers$ is the WZW level of the background and $h$ encodes the strength of the integrable interaction, and is a function of the RR charge and moduli values~\cite{OhlssonSax:2018hgc}. In the large worldsheet radius limit, low-lying physical states are made from a few magnons just like in flat space~\eqref{eq:flat-space-state}.\footnote{For more energetic states or at smaller worldsheet radius, this flat spacetime magnon picture is less useful.} 

Throughout this paper we will only be interested in $m=0$ states, since these are the only states that can give rise to extra protected multiplets.~\footnote{Extending the ABA to massive and mixed mass sectors is straightforward.} In the ABA a generic state is constructed in two steps. Firstly, one considers states made of $N_0$ massless magnons, each a \textit{fermionic} highest-weight state of  a $m=0$ $\alg{psu}(1|1)^4_{\ce}$ module with momentum $p_i$ (with $i=1,\dots,N_0)$\footnote{As in any super-module, the highest-weight state can be chosen to be bosonic or fermionic, depending on the choice of raising operators. In AdS/CFT, it is conventional to take the lightest $m>0$ excitations to be bosonic, in order to identify them with Hofman-Maldacena magnons. With this choice, the $m=0$ highest weight states then have to be fermionic.} Such states are physical if they satisfy the \textit{momentum carrying} BEs. Secondly, one acts with so-called $\aba{B}$ operators, which play the role of raising operators of the underlying Yangian algebra. They generate states whose constituent magnons include descendants of the $m=0$ $\alg{psu}(1|1)^4_{\ce}$ modules. These descendants should be thought of as further magnons on top of the $N_0$ magnons, as introduced in the famous nesting procedure (see~\cite{Slavnov:2019hdn} for a recent review). As such, they too carry momenta, which for physical states are constrained by \textit{auxiliary} BEs.
The algebra $\alg{psu}(1|1)^4_{\ce}$,  has four (fermionic) raising operators, but only two linear combinations of these act non-trivially on the \textit{short} magnon multiplets. We will label by $N_1$ and $N_3$ the number of times the corresponding $\aba{B}^I$ raising operators were used to produce a particular state and by $y_{I,k}$ the corresponding auxiliary momenta (with $I=1,3$ and $k=1,\dots,N_I$).

For example, taking each of the three magnons in~\eqref{eq:flat-space-state} to be massless,
with the fermionic magnon further chosen as highest-weight in $\alg{psu}(1|1)^4_{\ce}$\footnote{This would be done by suitably restricting the $\rep{8}_v$ and $\rep{8}_c$ labels.}, the state $\ket{v}$ has $N_0=3$ and $N_1=N_3=1$, since each of the bosons is an  $\alg{psu}(1|1)^4_{\ce}$ descendant. It will be convenient to write the state in terms of its Bethe roots as
\begin{equation}
\ket{v}\equiv\ket{\vec{p};y_1;y_3}
\end{equation}
where $\vec{p}=(p_1,p_2,p_3)$. If we had instead chosen the fermion to be lowest-weight, the state would still have $N_0=3$, but now it would have four auxiliary roots, with $N_1=N_3=2$. In flat space and in plane-wave backgrounds, the magnon S-matrix becomes the identity. As a result, the momentum carrying BEs reduce to the familiar integrality requirement on the momenta $p_i=\tfrac{n_i}{L}$ that follows from the periodicity of the worldsheet. The auxiliary BEs on the other hand trivialise completely to $1=1$. This is why auxiliary Bethe roots don't have a natural  interpretation in the familiar flat space or plane-wave setting, their multiplicity simply counting the number of times raising supercharges were used in any given state.

The  $\alg{psu}(1,1|2)^2$ charges of the ABA states can be determined from $N_i$, $p_i$ and an auxiliary integer parameter $L$ related to the length of the worldsheet by the gauge-fixing condition
\begin{equation}\label{modevsglobalT4}
\begin{aligned}
2D_{\sL}&= L+N_1+N_3-N_0+\sum_{i=1}^{N_0}
\sqrt{\k^2 p_i^2 +4 h^2 \sin^2\tfrac{p_i}{2}}
\\
2D_{\sR}&=L+\sum_{i=1}^{N_0}\sqrt{\k^2 p_i^2 +4 h^2 \sin^2\tfrac{p_i}{2}}
\\
2J_{\sL}&= L+N_1+N_3-N_0
\\
2J_{\sR}&=L,
\end{aligned}
\end{equation}
where $D$ and $J$ are, respectively, the $\alg{sl}(2)$ and $\alg{su}(2)$ Cartan generators of $\alg{psu}(1,1|2)_{\sL}\times \alg{psu}(1,1|2)_{\sR}$. The $\AdS_3$ backgrounds considered in this paper have \textit{small} $(4,4)$ superconformal symmetry. Protected states satisfy shortening conditions on both the left- and right-moving parts of the algebra
\begin{equation}
D_{\sL}=J_{\sL}\,,\qquad\qquad D_{\sR}=J_{\sR}\,.
\end{equation}
Such half-BPS multiplets are often written in the following notation
\begin{equation}\label{eq:sugra-mplet-notation}
\left(2D_{\sL}+1,2D_{\sR}+1\right)_{\sS}
\end{equation}
as, for example, reviewed in section 5.8 of~\cite{David:2002wn}. 

In this paper we use the ABA, to construct all such low-magnon number eigenvectors and show how they organise themselves into $\alg{psu}(1|1)^4_{\ce}$ multiplets. We then investigate the  $p_i\longrightarrow 0$ limit of these Bethe eigenvectors in order to identify the protected states. We show that in the strict $p_i=0$ limit all states become $\alg{psu}(1|1)^4_{\ce}$ singlets and the protected states are indeed made up of purely fermionic excitations, as anticipated in~\cite{Sax:2012jv}. It is interesting to note that these protected states come from both highest-weight and descendant states of the $p_i\neq0$ modules. The construction of the eigenvectors provided by the ABA allows us to straightforwardly generalise this analysis to $\AdS_3\times\Sphere^3\times \Torus^4/\Integers_k$, in other words the orbifold limit of $\AdS_3\times\Sphere^3\times\Kthree$,  and show that the protected spectrum is again the same as found using supergravity and dual $\CFT_2$ methods~\cite{deBoer:1998kjm}. 

This paper is organised as follows. In Section~\ref{sec:prot-spec} we review the protected spectrum of closed strings on  $\AdS_3\times\Sphere^3\times\Torus^4$ and 
$\AdS_3\times\Sphere^3\times\Kthree$ in its orbifold limits. In Section~\ref{sec:aba-sec} we review the general ABA construction and apply it to the $\AdS_3$ backgrounds of interest in the present paper. We end the section with a few examples of low-magnon number excitations and discuss their representation-theoretic properties. In Section~\ref{T4section} we apply the ABA methods to find the protected multiplets in $\AdS_3\times\Sphere^3\times\Torus^4$. We demonstrate that, in addition to taking the zero-momentum limit for momentum carrying roots, auxiliary roots for protected states also need to take special values given in equation~\eqref{eq:special-val-aux-root}. The short representations of the protected states follow from the number of momentum-carrying and auxiliary roots of a given multiplet as described in equation~\eqref{eq:prot-ket-de-boer}. In Section~\ref{sec:prot-stat-orb}, we use the explicit expressions for the protected eigenstates found in Section~\ref{T4section} to determine the spectrum of protected states in $\AdS_3\times\Sphere^3\times\Kthree$ orbifold backgrounds. In Section~\ref{sec:s3-and-orb-prot} we briefly apply our ABA analysis to the $\AdS_3\times\Sphere^3\times\Sphere^3\times\Sphere^1$ background and its $\Integers_2$ orbifold. Since none of the protected states involve auxiliary roots the analysis is much simpler, essentially following from our previous results~\cite{Baggio:2017kza}, taking into account the 
zero-momentum limit discussed in equation~\eqref{splusminusdef}. Finally, we present our conclusions and include four appendices to which the  technical details of some of our results are relegated. We also include a Mathematica notebook which generates explicit expressions for wavefunctions of states with up to four magnons and their descendants.

\section{Protected Spectrum of \texorpdfstring{$\AdS_3\times\Sphere^3\times\Torus^4$}{AdS3 x S3 x T4} and \texorpdfstring{$\AdS_3\times\Sphere^3\times\Kthree$}{AdS3 x S3 x K3}}
\label{sec:prot-spec}

In this section we briefly review the spectrum of protected closed string states on $\AdS_3\times\Sphere^3\times\Torus^4$ and $\AdS_3\times\Sphere^3\times\Kthree$, originally found from Kaluza-Klein supergravity reductions in~\cite{deBoer:1998kjm}. In the planar limit, holographic backgrounds with more supersymmetry, such as $\AdS_5\times \Sphere^5$ or $\AdS_4\times \CP^3$ have a single family of half-BPS BMN vacua, labelled by an integer $L$ equal to the angular momentum and conformal dimension of the dual operator \eg, $\tr(Z^L)$ in $\superN=4$ super-Yang-Mills theory. The corresponding string state is often written as
\begin{equation}\label{eq:bmn-vac}
\ket{Z^L}\,.
\end{equation}
The backgrounds considered in this paper have less supersymmetry and, as a result, have \textit{multiple} families of half-BPS vacua. Each family can be obtained by starting with the BMN vacuum~\eqref{eq:bmn-vac} and adding zero-momentum massless fermionic magnons~\cite{Sax:2012jv}. 

In the case of $\AdS_3\times\Sphere^3\times\Torus^4$, the protected multiplets organise themselves into a family of Hodge diamonds labelled by $L$, which we write following~\cite{Baggio:2017kza} as\footnote{$Z$, $\chi^{\dot{a}}$ and $\tilde{\chi}^{\dot{a}}$ respectively correspond to  $\Phi^{++}$, $\chi^{+\dot{a}}_{\sR}$ and $\chi^{+\dot{a}}_{\sL}$ in~\cite{Baggio:2017kza}.}
\begin{subequations}
\begin{gather}
 \ket{Z^L}
 \label{eq:0f}
 \\[4pt]
 \mathclap{\ket{Z^L\chi^{\dot{a}}}}
 \hspace{5cm}
 \mathclap{\ket{Z^L\tilde{\chi}^{\dot{a}}}}
 \label{eq:1f}
 \\[4pt]
 \mathclap{\epsilon_{\dot{a}\dot{b}}\ket{Z^L\chi^{\dot{a}}\chi^{\dot{b}}}}
 \hspace{5cm}
 \mathclap{\ket{ Z^L \chi^{\dot{a}}\tilde{\chi}^{\dot{b}} }}
 \hspace{5cm}
 \mathclap{\epsilon_{\dot{a}\dot{b}}\ket{Z^L\tilde{\chi}^{\dot{a}}\tilde{\chi}^{\dot{b}}}}
 \label{eq:2f}
 \\[4pt]
 \mathclap{\epsilon_{\dot{a}\dot{b}}\ket{Z^L\chi^{\dot{a}}\chi^{\dot{b}}\tilde{\chi}^{\dot{c}}}}
 \hspace{5cm}
 \mathclap{\epsilon_{\dot{a}\dot{b}}\ket{Z^L\chi^{\dot{a}}\tilde{\chi}^{\dot{b}}\tilde{\chi}^{\dot{c}} }}
 \label{eq:3f}
 \\[4pt]
\epsilon_{\dot{a}\dot{b}}\epsilon_{\dot{c}\dot{d}} \ket{Z^L\chi^{\dot{a}}\chi^{\dot{b}}
\tilde{\chi}^{\dot{c}}\tilde{\chi}^{\dot{d}}
 }
 \label{eq:4f}
\end{gather}
\end{subequations}
Above, the index $\dot{a}=\pm$  labels a $\rep{2}$ representation of the $\alg{su}(2)_\circ$ algebra that  is part of the rotations which act on the decompactified $\Torus^4$
\begin{equation}
\alg{so}(4)\sim\alg{su}(2)_\circ\oplus\alg{su}(2)_\bullet\,.
\end{equation}
Upon compactification, it is useful to continue labeling the zero winding and zero momentum excitations  in this way. In terms of the notation~\eqref{eq:sugra-mplet-notation}, these multiplets can  be written as
\begin{equation}\label{eq:t4-prot-spec-de-boer-notation}
  \begin{gathered}
   \left(L,L\right)_{\sS}^{\mathrlap{\rep{1}}}
    \\[2pt]
    \left(L,L+1\right)_{\sS}^{\mathrlap{\rep{2}}}
    \hspace{1cm}
    \left(L+1,L\right)_{\sS}^{\mathrlap{\rep{2}}}
    \\[2pt]
    \left(L,L+2\right)_{\sS}^{\mathrlap{\rep{1}}}
    \hspace{1cm}
    \left(L+1,L+1\right)_{\sS}^{\mathrlap{\rep{1}\oplus\rep{3}}}
    \hspace{1cm}
    \left(L+2,L\right)_{\sS}^{\mathrlap{\rep{1}}}
    \\[2pt]
    \left(L+1,L+2\right)_{\sS}^{\mathrlap{\rep{2}}}
    \hspace{1cm}
    \left(L+2,L+1\right)_{\sS}^{\mathrlap{\rep{2}}}
    \\[2pt]
    \left(L+2,L+2\right)_{\sS}^{\mathrlap{\rep{1}}}
  \end{gathered}
\end{equation}
with the superscripts denoting the $\alg{su}(2)_\circ$ representation and for compactness we have changed $L\rightarrow L-1$. In other words, for each $L$, the multiplets organise themselves into the Hodge diamond of $\Torus^4$  
\begin{equation}\label{eq:seed-t4}
  \begin{matrix}
    &&h^{0,0}&&\\
    &h^{1,0}&&h^{0,1}&\\ 
    h^{2,0}&&h^{1,1}&&h^{0,2}\\
    &h^{2,1}&&h^{1,2}&\\
    &&h^{2,2}&&
  \end{matrix}
  \quad=\quad
  \begin{matrix}
    &&1&&\\
    &2&&2&\\ 
    1&&4&&1\\
    &2&&2&\\
    &&1&&
  \end{matrix}
\end{equation}
The appearance of the cohomology of $T^4$ is to be expected given the conjectured connection of the protected operators of the dual $\CFT_2$ to the $\Sym^N$ orbifold~\cite{Maldacena:1997re}. It is well known that for such CFTs, the chiral ring of the theory is closely related to that of the Hodge diamond of the seed theory, which in this case is $\Torus^4$~\cite{Vafa:1994tf}. Because of this structure, in this paper we often express the protected spectrum of the theory in terms of the Hodge diamond of the seed theory.

Turning to the $\AdS_3\times\Sphere^3\times\Kthree$ background, we  blow down the  $\Kthree$ to an orbifold $\Torus/\Integers_n$ ($n=2,3,4,6$). The $\Integers_n$ orbifold action  acts only on the $\alg{su}(2)_\circ$ index as
\begin{equation}\label{orbifoldaction}
\left(\begin{array}{cc}\omega_n & 0 \\ 0 & \omega^{-1}_n \end{array}\right),
\end{equation}
where $\omega_n = e^{2\pi i/n}$. A simple way to see that this has to be the action of $\Integers_n$ is to note that  these orbifolds do not break any supersymmetry and the supercharges are singlets of  $\alg{su}(2)_\circ$, but doublets of $\alg{su}(2)_\bullet$. It is then clear that the states in equation~\eqref{eq:1f} and~\eqref{eq:3f} are projected out in all $\Integers_n$ orbifolds. The states in equations~\eqref{eq:0f} and~\eqref{eq:4f} on the other hand are always kept in all $\Integers_n$ orbifolds. All six states in equation~\eqref{eq:2f} are additionally kept in the $\Integers_2$ orbifold, while in $\Integers_n$ orbifolds with $n>2$ only the four states
\begin{equation}
\ket{Z^L\chi^{+}\chi^{-}}\,,\qquad 
\ket{Z^L\chi^{+}\tilde{\chi}^{-}}\,,\qquad 
\ket{Z^L\chi^{-}\tilde{\chi}^{+}}\,,\qquad 
\ket{Z^L\tilde{\chi}^{+}\tilde{\chi}^{-}}\,,
\end{equation}
are kept. In summary, for each $L$ in the $\Integers_2$ orbifold the untwisted sector Hodge numbers $h^{0,0}=h^{2,2}=h^{2,0}=h^{0,2}=1$ and $h^{1,1}=4$ while in 
$\Integers_{n>2}$ orbifolds we have from the untwisted sector the Hodge numbers $h^{0,0}=h^{2,2}=h^{2,0}=h^{0,2}=1$ and $h^{1,1}=2$.

The twisted sectors' Hodge numbers are well known (see for example~\cite{Walton:1987bu})
\begin{itemize}
\item $\Integers_2$ There are 16 identical twisted sectors. Each twisted sector is a blow-down of an $A_1$ two-cycle, giving a contribution to $h^{1,1}=16\times 1$ overall.
\item $\Integers_3$ There are 9 identical twisted sectors. Each twisted sector is a blow-down of an $A_2$ two cycle , giving a contribution to $h^{1,1}=9\times 2=18$ overall.
\item $\Integers_4$ There are 4 $A_3$ fixed points and 6 $A_1$ fixed points,  giving a contribution to $h^{1,1}=4\times 3+6\times 1=18$ overall.
\item $\Integers_6$ There is one $A_5$ fixed point and 4 $A_2$ fixed points and 5 $A_1$ fixed point, giving a contribution to $h^{1,1}=1\times 5+4\times 2+5\times 1=18$ overall.
\end{itemize}
In the $\Integers_2$ orbifold we therefore have the twisted sector Hodge numbers $h^{1,1}=16$, while in $\Integers_{n>2}$ orbifolds we have $h^{1,1}=18$. Adding the twisted and untwisted Hodge numbers in all cases gives the standard K3 Hodge diamond of the seed theory
\begin{equation}\label{eq:seed-k3}
  \begin{matrix}
    &&1&&\\
    &0&&0&\\ 
    1&&20&&1\\
    &0&&0&\\
    &&1&&
  \end{matrix}\,.
\end{equation}
The protected spectrum of the $\AdS_3\times\Sphere^3\times\Kthree$ theory is then a family of Hodge diamonds labelled by $L$
\begin{equation}\label{eq:k3-prot-spec-de-boer-notation}
  \begin{gathered}
   \left(L,L\right)_{\sS}
    \\[2pt]
    \mbox{\O}
    \hspace{3.5cm}
     \mbox{\O}
    \\[2pt]
    \left(L,L+2\right)_{\sS}
    \hspace{1cm}
    \left(L+1,L+1\right)_{\sS}^{\oplus 20}
    \hspace{1cm}
    \left(L+2,L\right)_{\sS}
    \\[2pt]
    \mbox{\O}
    \hspace{3.5cm}
    \mbox{\O}
    \\[2pt]
    \left(L+2,L+2\right)_{\sS}
  \end{gathered}
\end{equation}
Since the orbifold action breaks $\alg{su}(2)_\circ$, we have removed the representation-theoretic superscript compared to equation~\eqref{eq:t4-prot-spec-de-boer-notation}, listing instead multiplicity in the superscript where it is non-trivial.

\section{Algebraic Bethe ansatz for \texorpdfstring{$\AdS_3 \times \Sphere^3 \times \Torus^4$}{AdS3 x S3 x T4}}
\label{sec:aba-sec}

In this section we present the Algebraic Bethe ansatz for strings on $\AdS_3 \times \Sphere^3 \times \Torus^4$. We begin by reviewing the representation-theory and R matrix of world-sheet excitations, focusing on the massless sector in Section~\ref{sec:representations}. We then summarise the general Bethe ansatz prescription for eigenstates in terms of the transfer matrix in Section~\ref{sec:ba-estates} and ABA construction of eigenstates using the monodromy matrix in Section~\ref{sec:ABA}. We end with some examples of low-magnon states in the massless sector of $\AdS_3 \times \Sphere^3 \times \Torus^4$ in Section~\ref{sec:unprot-states}.

\subsection{Representations of \texorpdfstring{$\algPSU(1|1)^2_{\ce}$}{psu(1|1)2} and \texorpdfstring{$\algPSU(1|1)^4_{\ce}$}{psu(1|1)4}}
\label{sec:representations}

The supercharges of the $\alg{psu}(1|1)^2_{\ce}$ algebra satisfy the commutation relations
\begin{equation}\label{eq:psu112-comm-rel}
  \acomm{\gen{Q}_{\sL}}{\gen{S}_{\sL}} = \gen{H}_{\sL} , \qquad
  \acomm{\gen{Q}_{\sR}}{\gen{S}_{\sR}} = \gen{H}_{\sR}, \qquad
  \acomm{\gen{Q}_{\sL}}{\gen{Q}_{\sR}} = \gen{C} , \qquad
  \acomm{\gen{S}_{\sL}}{\gen{S}_{\sR}} = \bar{\gen{C}} ,
\end{equation}
with $\gen{H}_{\sL}$, $\gen{H}_{\sR}$, $\gen{C}$, $\bar{\gen{C}}$ central elements. The algebra furthermore comes equipped with a non-trivial coproduct of the form
\begin{equation}\label{eq:coprod}
  \begin{aligned}
    \Delta(\gen{Q}_{\sL}) &= \gen{Q}_{\sL} \otimes 1 + e^{+\frac{i}{2}\gen{P}} \otimes \gen{Q}_{\sL}, \qquad
    \Delta(\gen{S}_{\sL}) &= \gen{S}_{\sL} \otimes 1 + e^{-\frac{i}{2}\gen{P}} \otimes \gen{S}_{\sL}, \\
    \Delta(\gen{Q}_{\sR}) &= \gen{Q}_{\sR} \otimes 1 + e^{+\frac{i}{2}\gen{P}} \otimes \gen{Q}_{\sR}, \qquad
    \Delta(\gen{S}_{\sR}) &= \gen{S}_{\sR} \otimes 1 + e^{-\frac{i}{2}\gen{P}} \otimes \gen{S}_{\sR},
  \end{aligned}
\end{equation}
where $\gen{P}$ is the world-sheet momentum.\footnote{The coproduct of the momentum operator is given by
  \begin{equation*}
    \Delta(e^{i\gen{P}}) = e^{i\gen{P}} \otimes e^{i\gen{P}} .
  \end{equation*}
  The coproduct of the central charges $\gen{C}$ and $\bar{\gen{C}}$ is also non-trivial, and follows from the commutation relations~\eqref{eq:psu112-comm-rel} together with equation~\eqref{eq:coprod}.}

Throughout this paper we will use two families of short representations of $\alg{psu}(1|1)^2_{\ce}$ which we denote by $\rho_{\sL}$ and $\tilde{\rho}_{\sL}$. The short representations $\rho_{\sL}$, labeled by  momentum $p$, act on the graded vector space $ \mathcal{V}_{\rho_{\sL}}=\{\ket{\phi_p},\ket{\psi_p}\}$ as
\begin{equation}\label{eq:rep-L}
  \begin{aligned}
    \gen{Q}_{\sL}\ket{\phi_p} &= \eta_p \ket{\psi_p},\qquad &
    \gen{Q}_{\sR}\ket{\psi_p} &= -\frac{\eta_p}{x_p^-} e^{-ip/2}\ket{\phi_p},\\
    \gen{S}_{\sL}\ket{\psi_p} &= \eta_p e^{-ip/2}\ket{\phi_p},\qquad &
    \gen{S}_{\sR}\ket{\phi_p} &= -\frac{\eta_p}{x_p^+} \ket{\psi_p},
  \end{aligned}
\end{equation}
where 
\begin{equation}\label{def:etaL}
  \eta_p=e^{\frac{ip}{4}}\sqrt{\tfrac{ih}{2}(x_{p}^--x_{p}^+)} .
\end{equation}
The Zhukovski variables $x_{p}^\pm$ are related to the momentum $p$ through the relations
\begin{equation}\label{eq:momentum-shortening}
  \frac{x_p^+}{x_p^-} = e^{ip} , \qquad
  x_p^+ + \frac{1}{x_p^+} - x_p^- - \frac{1}{x_p^-} = \frac{2i(|m|+\k p)}{h} .
\end{equation}
which are solved in the physical region by
\begin{equation}\label{LZhukovskitomommixed}
  x_p^{\pm}=\frac{(|m|+\k p)+\sqrt{(|m|+\k p)^2+4h^2\sin^2{\frac{p}{2}}}}{2h\sin{\frac{p}{2}}}e^{\pm \frac{i p}{2}},\qquad \k=\frac{k}{2\pi}.
\end{equation}
Above, $h$ is the integrable-interaction coupling constant, which is a function of the moduli,  $k\in\Integers$ is the NS-NS charge of the background and $m$ is the mass of the excitations.  The representations $\tilde{\rho}_{\sL}$ act on $\mathcal{V}_{\tilde{\rho}_{\sL}}=\{\ket{\tilde{\psi}_p},\ket{\tilde{\phi}_p}\}$ as
\begin{equation}\label{eq:rep-L-tilde}
  \begin{aligned}
    \gen{Q}_{\sL}\ket{\tilde{\psi}_p} &= \eta_p \ket{\tilde{\phi}_p}, \qquad &
    \gen{Q}_{\sR}\ket{\tilde{\phi}_p} &= -\frac{\eta_p}{x_p^-} e^{-ip/2}\ket{\tilde{\psi}_p},\\
    \gen{S}_{\sL}\ket{\tilde{\phi}_p} &= \eta_p e^{-ip/2}\ket{\tilde{\psi}_p}, &
    \gen{S}_{\sR}\ket{\tilde{\psi}_p} &= -\frac{\eta_p}{x_p^+} \ket{\tilde{\phi}_p},
  \end{aligned}
\end{equation}
and are obtained from $\rho_{\sL}$  by swapping the grading of the two excitations. 

The exact $\alg{psu}(1|1)^2_{\ce}$-invariant R matrices for these representations are
\begin{equation}\label{eq:RLLmixed-RtLtLmixed}
  R^{\sL \sL}(p,q)=\begin{pmatrix}
    A_{pq}&0&0&0\\
    0&B_{pq}&E_{pq}&0\\
    0&C_{pq}&D_{pq}&0\\
    0&0&0&-F_{pq}
  \end{pmatrix},
  \qquad
  R^{\tilde{\sL}\tilde{\sL}}(p,q)=\begin{pmatrix}
    A_{pq}&0&0&0\\
    0&B_{pq}&-E_{pq}&0\\
    0&-C_{pq}&D_{pq}&0\\
    0&0&0&-F_{pq}
  \end{pmatrix},
\end{equation} 
where
\begin{equation}\label{ABCDEFcoeffsLrep}
  \begin{aligned}
    A_{pq} &= 1, &
    \quad
    B_{pq} &= \phantom{-}\left( \frac{x^-_{p}}{x^+_{p}}\right)^{1/2} \frac{x^+_{p}-x^+_{q}}{x^-_{p}-x^+_{q}}, \\
    C_{pq} &= \left( \frac{x^-_{p}}{x^+_{p}} \frac{x^+_{q}}{x^-_{q}}\right)^{1/2} \frac{x^-_{q}-x^+_{q}}{x^-_{p}-x^+_{q}} \frac{\eta_{p}}{\eta_{q}}, 
    \quad &
    D_{pq} &= \phantom{-}\left(\frac{x^+_{q}}{x^-_{q}}\right)^{1/2}  \frac{x^-_{p}-x^-_{q}}{x^-_{p}-x^+_{q}}, \\
    E_{pq} &= \frac{x^-_{p}-x^+_{p}}{x^-_{p}-x^+_{q}} \frac{\eta_{q}}{\eta_{p}}, 
    \quad &
    F_{pq} &= - \left(\frac{x^-_{p}}{x^+_{p}} \frac{x^+_{q}}{x^-_{q}}\right)^{1/2} \frac{x^+_{p}-x^-_{q}}{x^-_{p}-x^+_{q}}.
  \end{aligned}
\end{equation}

The $\alg{psu}(1|1)^4_{\ce}$ algebra is a tensor product of  two commuting copies of $\alg{psu}(1|1)^2_{\ce}$, and throughout this paper we will use the short representation 
$\rho_{\alg{psu}(1|1)^4}$
\begin{equation}\label{tensorrep}
    \rho_{\alg{psu}(1|1)^4}=\rho_{\sL}\otimes \tilde{\rho}_{\sL},
\end{equation}
which acts on  $ \mathcal{V}_p\equiv \mathcal{V}_{\rho_{\sL}}\otimes\mathcal{V}_{\tilde{\rho}_{\sL}}$ whose basis elements we write as
\begin{equation}\label{basissu114}
\ket{\chi}\equiv\ket{\phi}\ket{\tilde{\psi}},\qquad \ket{T^2}\equiv\ket{\phi}\ket{\tilde{\phi}},\qquad \ket{T^1}\equiv\ket{\psi}\ket{\tilde{\psi}},\qquad \ket{\tilde{\chi}}\equiv\ket{\psi}\ket{\tilde{\phi}},
\end{equation}
in order to emphasize their connection to the massless worldsheet excitations that appear in the gauge-fixed Lagrangian~\cite{Borsato:2014exa,Borsato:2014hja}.\footnote{\label{ft:su2-circ} The massless worldsheet excitations additionally all transform in the fundamental representation of $\alg{su}(2)_\circ$ and in total there are four bosons and four fermions in the massless sector. However, $\alg{su}(2)_\circ$ does not enter the Bethe equations or the R-matrix and we will drop the corresponding index (denoted by ${}^a$ in~\cite{Borsato:2014exa,Borsato:2014hja}) in this section to declutter the notation. This multiplicity will be important when counting the number of protected states and so we will re-instate the  $\alg{su}(2)_\circ$ index in Section~\ref{T4section}.} In the basis~\eqref{basissu114}, the supercharges act as\footnote{This matches equation~(3.35) of \cite{Lloyd:2014bsa} with the phase $\xi=0$.}
\begin{equation}\label{superchargerepsu114}
  \begin{aligned}
    \gen{Q}_{\sL}^1=-x_p^+ \gen{S}_{\sR}^1 &= \eta_p^L\begin{pmatrix}
      0&0&0&0\\
      0&0&0&0\\
      1&0&0&0\\
      0&1&0&0
    \end{pmatrix},
    \qquad &
    \gen{S}_{\sL}^1=-x_p^-\gen{Q}_{\sR}^1&=e^{-\frac{ip}{2}}\eta_p^L\begin{pmatrix}
      0&0&1&0\\
      0&0&0&1\\
      0&0&0&0\\
      0&0&0&0
    \end{pmatrix},
    \\[8pt]
    \gen{Q}_{\sL}^2=-x_p^+\gen{S}_{\sR}^2 &= \eta_p^L\begin{pmatrix}
      0&0&0&0\\
      1&0&0&0\\
      0&0&0&0\\
      0&0&-1&0
    \end{pmatrix},
    \qquad &
    \gen{S}_{\sL}^2=-x_p^-\gen{Q}_{\sR}^2&=e^{-\frac{ip}{2}}\eta_p^L\begin{pmatrix}
      0&1&0&0\\
      0&0&0&0\\
      0&0&0&-1\\
      0&0&0&0
    \end{pmatrix}\,.
  \end{aligned}
\end{equation}
The $\alg{psu}(1|1)^4$ R-matrix is the graded tensor product of the $\alg{psu}(1|1)^2$ R-matrices 
\begin{equation}\label{Rmatrixsu114}
    R_{\alg{psu}(1|1)^4}= R^{\sL\sL}_{\alg{psu}(1|1)^2} \otimes R^{\tilde\sL\tilde\sL}_{\alg{psu}(1|1)^2}.
\end{equation}
For the special case of pure RR background ($k=0$) and massless excitations the above R-matrices can be written in a difference-form using relativistic $\gamma$ variables~\cite{Fontanella:2019baq}. We discuss the pure RR results in those variables in Appendix~\ref{appendix:pureRR}. In the remainder of this section, we use the R-matrix in equation~\eqref{Rmatrixsu114} to find the string spectrum using the Bethe Ansatz.

\subsection{Bethe ansatz eigenstates}
\label{sec:ba-estates}

The spectrum of $\AdS_3\times \Sphere^3\times \Torus^4$ can be obtained through a Bethe ansatz. In this paper we are mainly interested in protected states, which are built by adding fermionic zero modes on top of BMN ground states. To understand how such protected states are obtained we first review how to construct generic wave functions using the Bethe ansatz before specialising to the protected states.\footnote{For more details in the $\AdS_3/\CFT_2$ context see for example reference~\cite{Borsato:2016hud}.}

The basic building block of the Bethe ansatz for the wave functions are asymptotic states with well-separated excitations carrying momenta $p_i$. We write such a state as
\begin{equation}
  \ket{ \Phi^{\alpha_1}_{p_1} \dotsb \Phi^{\alpha_N}_{p_N} } ,
\end{equation}
where $\Phi^{\alpha}$ represents any type of excitation. Note that there is an implied ordering of the excitations, which can be highlighted by going to a position basis
\begin{equation}\label{eq:plane-wave-exp}
  \ket{\Phi^{\alpha_1}_{p_1} \dotsb \Phi^{\alpha_N}_{p_N}}
  =
  \int\limits_{\mathclap{\sigma_1 \ll \dotsb \ll \sigma_N}} \;d\sigma_1 \dotsb d\sigma_N \, e^{i(p_1 \sigma_1 + \dotsb + p_N \sigma_N)} \ket{\Phi^{\alpha_1}(\sigma_1) \dotsb \Phi^{\alpha_N}(\sigma_N) } .
\end{equation}
In order to build a full eigenstate from such asymptotic states we make an ansatz where we sum over all orderings of the excitations\footnote{This ansatz does not give a complete description of the eigenstate, since we only consider excitations that are well separated, and thus ignore any region where the excitations come close enough to each other to interact. As we will see below, interactions are taken into account when we glue the different regions together using the two-particle S matrix.}
\begin{equation}\label{eq:Psi-as-permutations}
  \ket{\Psi}
  =
  \sum_{\tau \in S_N} \psi^{\tau}_{\alpha_1\dotsb\alpha_N}(\vec{p}) \ket{ \Phi^{\alpha_1}_{p_{\tau(1)}} \dotsb \Phi^{\alpha_N}_{p_{\tau(N)}} } ,
\end{equation}
where we use $\vec{p}$ as a shorthand for $p_1,\dotsc,p_N$.
The partial wave functions $\psi^{\tau}_{\alpha_1\dotsb\alpha_N}(\vec{p})$ are related by the two-particle S matrix, which acts on two neighbouring excitations as\footnote{The S matrix is related to the R matrix by the relation $\Rmat = \Pi \circ \Smat$, where $\Pi$ is a (graded) permutation.}
\begin{equation}
  \Smat_{12} \ket{\Phi_{p_1}^{\alpha_1} \Phi_{p_2}^{\alpha_2}} = S^{\alpha_1 \alpha_2}_{\beta_1 \beta_2}(p_1,p_2) \ket{\Phi_{p_2}^{\beta_2} \Phi_{p_1}^{\beta_1}} .
\end{equation}
For the wave function corresponding to a permutation $\tau$ we have relations of the form
\begin{equation}\label{eq:psi-S-psi}
  \psi^{\tau}_{\alpha_1 \dotsb \alpha_{i+1} \alpha_i \dotsb \alpha_N} (\vec{p})
  =
  S^{\beta_i \beta_{i+1}}_{\alpha_i \alpha_{i+1}} (p_i, p_{i+1}) \psi^{(i\,i+1) \circ \tau}_{\alpha_1 \dotsb \beta_i \beta_{i+1} \dotsb \alpha_N} (\vec{p}) ,
\end{equation}
where the wave function on the right-hand side corresponds to the composition of the transposition $(i\,i+1)$ and $\tau$. Since any permutation can be factored into a series of transpositions of nearest neighbours all of the partial wave functions can be related to each other in this way. Such a decomposition is in general not unique, but if the two-particle S matrix satisfies the Yang-Baxter equation, all possible decompositions of a permutation lead to the same wave function. This construction furthermore assumes integrability, which implies that any interaction can be separated into distinct two-particle interactions.

As an example of the above construction, let us consider the simplest case of a model with a single type of bosonic excitation $\phi$, and a state with two excitations ($N=2$) with momenta $p_1$ and $p_2$, with $p_1 > p_2$. We then have two possible orderings: the incoming wavepacket
\begin{equation}
  \ket{\phi_{p_1} \phi_{p_2}}
  =
  \int\limits_{\mathclap{\sigma_1 \ll \sigma_2}} \;d\sigma_1 d\sigma_2 \, e^{i(p_1 \sigma_1 + p_2 \sigma_2)} \ket{\phi(\sigma_1) \phi(\sigma_2) } ,
\end{equation}
and the outgoing wavepacket
\begin{equation}
  \ket{\phi_{p_2} \phi_{p_1}}
  =
  \int\limits_{\mathclap{\sigma_1 \ll \sigma_2}} \;d\sigma_1 d\sigma_2 \, e^{i(p_2 \sigma_1 + p_1 \sigma_2)} \ket{\phi(\sigma_1) \phi(\sigma_2) } .
\end{equation}
The state in equation~\eqref{eq:Psi-as-permutations} is now a sum over two terms
\begin{equation}
  \ket{\Psi_2}
  =
  \psi^{e}(p_1,p_2) \ket{\phi_{p_1} \phi_{p_2}} + \psi^{(12)}(p_1,p_2) \ket{\phi_{p_2} \phi_{p_1}},
\end{equation}
where $e$ denotes the identity permutation. As in equation~\eqref{eq:psi-S-psi} the two wavefunctions above are related by
\begin{equation}
  \psi^{(12)}(p_1,p_2) = S(p_1,p_2) \psi^{e}(p_1,p_2) ,
\end{equation}
where the S matrix is now replaced by the scattering phase $S(p_1,p_2)$. Using this, the state $\ket{\Psi}$ is, up to an overall factor, just the sum over the incoming and outgoing components
\begin{equation}
  \ket{\Psi_2}
  =
  \psi^{e}(p_1,p_2) \Bigl( \ket{\phi_{p_1} \phi_{p_2}} + S(p_1,p_2) \ket{\phi_{p_2} \phi_{p_1}} \Bigr) .
\end{equation}

The above wave functions describe the system on an infinite line. For a closed string we want to impose periodic boundary conditions on a circle of some length $L$. To do this it is useful to introduce the \emph{full} spatial wave function $\Psi_{\alpha_1 \dotsb \alpha_N}(\sigma_1,\dotsc,\sigma_N)$ in terms of which the state $\ket{\Psi}$ can be written as
\begin{equation}\label{eq:Psi-as-Psi}
  \ket{\Psi}
  =
  \int\limits_{\mathclap{\sigma_1 \ll \dotsb \ll \sigma_N}} \;d\sigma_1 \dotsb d\sigma_N \, \Psi_{\alpha_1 \dotsb \alpha_N}(\sigma_1,\dotsc,\sigma_N) \ket{\Phi^{\alpha_1}(\sigma_1) \dotsb \Phi^{\alpha_N}(\sigma_N)} .
\end{equation}
Periodic boundary conditions then imply that
\begin{equation}
  \Psi_{\alpha_1 \dotsb \alpha_{N-1} \alpha_N}(\sigma_1,\dotsc,\sigma_{N-1},\sigma_N) = \Psi_{\alpha_2 \dotsb \alpha_N \alpha_1}(\sigma_2,\dotsc,\sigma_N,\sigma_1 + L) .
\end{equation}

Expanding out the full wave function using equations~\eqref{eq:plane-wave-exp} and~\eqref{eq:Psi-as-Psi}, and repeatedly using relation~\eqref{eq:psi-S-psi} so that all the partial wave functions are written in terms of the wave function corresponding to the same permutation $\tau$ we find that demanding periodicity gives a set of equations of the form
\begin{equation}\label{eq:BE-psi-T}
  \psi^{\tau}_{a_1 \dotsb a_N}(\vec{p})
  =
  -e^{i p_k L} \aba{T}^{b_1\dotsb b_N}_{a_1 \dotsb a_N}(p_k | \vec{p})
  \psi^{\tau}_{b_1\dotsb b_N}(\vec{p}) ,
  \quad
  k = 1,\dotsc N,
\end{equation}
where the matrix $\aba{T}$ is the \emph{transfer matrix}, which we will discuss in more detail below.

For the simple $N=2$ example with a single flavour of excitations, the full state can be written as
\begin{equation}
  \ket{\Psi_2}
  =
  \int\limits_{\mathclap{\sigma_1 \ll \sigma_2}} \;d\sigma_1 d\sigma_2 \,
  \psi^{e}(p_1,p_2) \, \Bigl( e^{i(p_1\sigma_1 + p_2\sigma_2)} + S(p_1,p_2) e^{i(p_2\sigma_1 + p_1\sigma_2)} \Bigr) \ket{\phi(\sigma_1)\phi(\sigma_2)} ,
\end{equation}
from which we identify the wave function
\begin{equation}\label{eq:total-2mag-state}
  \Psi(\sigma_1,\sigma_2) = \psi^{e}(p_1,p_2) \, \Bigl( e^{i(p_1\sigma_1 + p_2\sigma_2)} + S(p_1,p_2) e^{i(p_2\sigma_1 + p_1\sigma_2)} \Bigr) .
\end{equation}
Periodic boundary conditions now gives us that
\begin{equation}
  \Psi(\sigma_1, \sigma_2) = \Psi(\sigma_2,\sigma_1 + L) ,
\end{equation}
which leads to the two conditions
\begin{equation}
  \psi^{e}(p_1,p_2) = e^{i p_1 L} S(p_1,p_2) \psi^{e}(p_1,p_2) , \qquad
  \psi^{(12)}(p_1,p_2) = e^{i p_2 L} S(p_2,p_1) \psi^{(12)}(p_1,p_2) .
\end{equation}
Assuming that $S(p,p) = -1\,,$ this can be written as the constraint\footnote{In this simple example there is only a single flavour and hence $\aba{T}(p_1|p_1,p_2)$ and $\aba{T}(p_2|p_1,p_2)$ are just numbers
  \begin{equation*}
    \begin{gathered}
      \aba{T}(p_1|p_1,p_2) = S(p_1,p_1) S(p_1,p_2) = -S(p_1,p_2) , \\
      \aba{T}(p_2|p_1,p_2) = S(p_2,p_1) S(p_2,p_2) = -S(p_2,p_1) = -\frac{1}{S(p_1,p_2)} .
    \end{gathered}
  \end{equation*}
}
\begin{equation}
  1 + e^{ip_kL} \prod_{j = 1}^{2} S(p_k,p_j) = 0,
\end{equation}
which for $k=1,2$ gives the quantisation conditions for the two momenta $p_1$ and $p_2$.

To solve the equations~\eqref{eq:BE-psi-T} we first need to diagonalise $\aba{T}$. Once we have found the eigenvalues and eigenvectors of $\aba{T}$, equation~\eqref{eq:BE-psi-T} gives quantisation conditions on the momenta $p_j$. For a given set of momenta solving these Bethe equations, we can finally reconstruct the full eigenstate by summing over all permutations of the momenta with coefficient obtained from equation~\eqref{eq:psi-S-psi}.

\subsection{The algebraic Bethe ansatz}\label{sec:ABA}

There are many ways to diagonalise $\aba{T}$. Here we will mainly focus on the algebraic Bethe ansatz (ABA) construction, while in Appendix~\ref{app:CBA} we present a complementaty coordinate Bethe ansatz construction.
In the ABA in order to diagonalize $\aba{T}$, one introduces a more general object, the \emph{monodromy matrix} $\aba{M}$. To construct $\aba{M}$ we consider a state with some number of physical excitations, and add to it an additional excitation, referred to as an \emph{auxiliary} excitation. The auxiliary excitation starts out to the left of all the excitations, and we let it scatter once with physical excitation until it sits to the right of all of them.

For concreteness we will now consider states consisting of massless $\AdS_3 \times \Sphere^3 \times \Torus^4$ excitations. We take the auxiliary excitation to sit in a $\rho_{\alg{psu}(1|1)^4}$ representation\footnote{\label{ft:massive-aux}The transfer matrix obtained for any short representation has the same set of eigenvectors, and hence the same final physical spectrum. It is convenient to pick the auxiliary excitations to be in a \textit{massive} representation even if all physical excitations are massless. With this choice, the descendent states arise from auxiliary roots in the conventional fashion~\cite{Beisert:2005fw}.} of $\alg{psu}(1|1)^4_{\ce}$. Viewed as a linear map on the auxiliary space, the monodromy matrix $\aba{M}$ can then be written as a $4 \times 4$ matrix whose entries are operators acting on the physical excitations. Since the $\alg{psu}(1|1)^4_{\ce}$ algebra is a direct product of two copies of the $\alg{psu}(1|1)^2_{\ce}$ algebra, the $\alg{psu}(1|1)^4_{\ce}$ monodromy matrix can be written as a product of two $2 \times 2$ $\alg{psu}(1|1)^2_{\ce}$ monodromy matrices.\footnote{As discussed in section~\ref{sec:representations} the massless excitations transform in the $\rho_{\sL} \otimes \rho_{\tilde{\sL}}$ representation of $\algPSU(1|1)^4_{\ce}$, and hence in slightly different representations under the two copies of $\algPSU(1|1)^2_{\ce}$. It is therefore also convenient to also pick the representations of two auxiliary spaces 1 and 3 to be massive versions of the $\rho_{\sL}$ and $\rho_{\tilde{\sL}}$. The spectra of the two transfer matrices will be almost the same up to some fermion minus signs.}  We denote the components of this smaller matrix by
\begin{equation}\label{monodromysu112comp}
  \aba{M}^I(p_0) =
  \begin{pmatrix}
    \aba{A}^I(p_0) & \aba{B}^I(p_0) \\
    \aba{C}^I(p_0) & \aba{D}^I(p_0)
  \end{pmatrix} \,,
\end{equation}
where the index $I=1,3$ labels the two copies of $\alg{psu}(1|1)^2_{\ce}$ and we have indicated the dependence on the momentum $p_0$ of the auxiliary excitations.
The transfer matrix is the supertrace of the monodromy matrix over the auxiliary space\footnote{Here, we have suppressed the dependence of $\aba{T}$ on the momenta $p_1,\dots,p_N$ for brevity.}
\begin{equation}
  \aba{T}^I(p_0) = \str_0 \aba{M}^I(p_0) \,.
\end{equation}
Using the Yang-Baxter equation one can show that the transfer matrix commutes with itself for any value of the auxiliary momentum
\begin{equation}
  \comm{\aba{T}^I(p_0)}{\aba{T}^I(p_0')} = 0 \,.
\end{equation}
In the Bethe equation~\eqref{eq:BE-psi-T} we need the eigenvalues of $\aba{T}^I(p_0)$ when the auxiliary momentum coincides with one of the physical momenta $p_i$, but because of the above equation we can diagonalise $\aba{T}$ for all auxiliary momenta simultaneously.

The simplest eigenvectors of $\aba{T}$ are states made of $N_0$ $\alg{psu}(1|1)^4_{\ce}$ highest-weight excitations $\chi$, since these all scatter diagonally
\begin{equation}\label{eq:n0-mag-pseud-state}
 \ket{\chi_{p_1} \dotsb \chi_{p_{N_0}}} \,.
\end{equation}
Other eigenvectors are obtained by acting with  $N_I$ $\aba{B}^I$ operators with $I=1,3$, which act as raising operators in the algebraic Bethe Ansatz. For example, given the above reference state, we  obtain a new state with $N_0-1$ $\chi$ excitations and one $T^1$, by acting with a single $\aba{B}^1$ operator
\begin{equation}
 \aba{B}^1(y) \ket{\chi_{p_1} \dotsb \chi_{p_{N_0}}} \,,
\end{equation}
where now the argument of $\aba{B}^1$ is the auxiliary Zhukovski parameter $y \equiv x^-_{q}$ for the auxiliary momentum $q$.\footnote{\label{fn:B-normalisation}%
  The S matrix, and hence also the monodromy matrix and transfer matrix, depend on the Zhukovski parameters $x^{\pm}_q$ of the auxiliary excitation. However, the state generated by the action of $\aba{B}^1(y)$ can always be written in a form such that the $x^+_q$ dependence enters only in the overall normalisation of the state, while the relative coefficients only depend explicitly on $x^-_q$. For a state of the form
  \begin{equation*}
    \aba{B}^{1}(y_{1}) \dotsb \aba{B}^{1}(y_{{N_1}}) \ket{\chi_{p_1} \dotsb \chi_{p_{N_0}}} \,,
  \end{equation*}
  the $x^+_{q_i}$ dependence is captured by a factor of the form
  \begin{equation*}
    \prod_{i=1}^{N_1} \sqrt{\frac{x^-_{q_i}}{x^+_{q_i}}} \frac{\eta_{q_i}}{\prod_{j<i} D_{q_i q_j}} \,,
  \end{equation*}
  with $\eta$ and $D$ defined in equations~\eqref{def:etaL} and~\eqref{ABCDEFcoeffsLrep}, respectively.
  Since the normalisation of states is not important in our discussion, we will write the argument of $\aba{B}$ as $y = x^-_q$ only, and drop the above normalization factor when writing out explicit expressions for the action of $\aba{B}$.}
The other eigenstates can be built by acting with more creation operators
\begin{equation}\label{ABAsu114}
\ket{\vec{p};\vec{y_1};\vec{y_3}}\equiv \aba{B}^{1}(y_{1,1}) \dotsb \aba{B}^{1}(y_{1,{N_1}}) \aba{B}^{3}(y_{3,1}) \dotsb \aba{B}^{3}(y_{3,{N_3}}) \ket{\chi_{p_1} \dotsb \chi_{p_{N_0}}} \,.
\end{equation}
where $\vec{p}=\{p_1,\dots,p_{N_0}\}$ and $\vec{y_{I}}=\{y_{I,1},\dots,y_{I,N_I}\}$
Acting with the two transfer matrices on such a state gives
\begin{equation}\label{eq:ABA-eigenstate-X}
  \aba{T}^1(p_0) \aba{T}^3(p_0) \ket{\vec{p};\vec{y_1};\vec{y_3}}
  =
  \Lambda^1(p_0|\vec{y}_{1}|\vec{p})
  \Lambda^3(p_0|\vec{y}_{3}|\vec{p})
  \ket{\vec{p};\vec{y_1};\vec{y_3}}
  +
  \ket{X} \,.
\end{equation}
The \emph{unwanted term} $\ket{X}$ is a linear combination of states with one less auxiliary root than $ \ket{\vec{p};\vec{y_1};\vec{y_3}}$ on which a single $\aba{B}^J(p_0)$ acts. These vanish provided the  $y_{I,k}$ satisfy the auxiliary Bethe equations
\begin{equation}\label{eq:BE-aux}
  1 = \prod_{j=1}^{N_0} \sqrt{\frac{x_j^+}{x_j^-}} \frac{y_{I,k} - x_j^-}{y_{I,k} - x_j^+} , \qquad k = 1,\dotsc,N_I, \qquad I = 1,3 \,.
\end{equation}
The eigenvalue in equation~\eqref{eq:ABA-eigenstate-X} is given by
\begin{equation}
  \Lambda^1(p_0|\vec{y}_{1}|\vec{p}) \Lambda^3(p_0|\vec{y}_{3}|\vec{p})
  =
  -\left(1-\prod_{i=1}^{N_0}\sqrt{\frac{x^+_{i}}{x^-_{i}}}  \frac{x_0^--x^-_{i}}{x_0^--x^+_{i}}\right)^2
  \prod_{I=1,3} \prod_{j=1}^{N_I}\frac{x_{0}^+-y_{I,j}^-}{x_0^--y_{1,j}^-}\sqrt{\frac{x_0^-}{x_0^+}}\,.
\end{equation}
In the Bethe equation~\eqref{eq:BE-psi-T} $\Lambda^I$ is evaluated for $p_0$ equal to one of the physical momenta $p_i$, in which case the above expression reduces to
\begin{equation}
  \Lambda^1(p_k|\vec{y}_{1}|\vec{p}) \Lambda^3(p_k|\vec{y}_{3}|\vec{p})
  =
  -\!\prod_{I=1,3} \prod_{j=1}^{N_I}\frac{x_k^+-y_{I,j}^-}{x_k^--y_{1,j}^-}\sqrt{\frac{x_k^-}{x_k^+}} \,,
\end{equation}
since now $x_0^-=x_i^-$ for one of the $i$. Using the normalisation in the S matrix of~\cite{Borsato:2016xns}, this leads to the momentum carrying Bethe equation
\begin{equation}\label{eq:BE-momentum-carrying}
  \left(\frac{x_k^+}{x_k^-}\right)^L=\prod_{\substack{j=1\\ j \neq k}}^{N_0} \sqrt{\frac{x_k^-}{x_k^+}\frac{x_j^+}{x_j^-}} \frac{x_k^+-x_j^-}{x_k^--x_j^+} (\sigma_{kj}^{\circ\circ})^2 \prod_{j=1}^{N_1} \sqrt{\frac{x_k^+}{x_k^-}} \frac{x_k^--y_{1,j}}{x_k^+-y_{1,j}}\prod_{j=1}^{N_3} \sqrt{\frac{x_k^+}{x_k^-}} \frac{x_k^--y_{3,j}}{x_k^+-y_{3,j}}\,.
\end{equation}
Above, $\sigma_{kj}^{\circ\circ}$ is the massless dressing factor~\cite{Borsato:2016kbm}.

\subsection{Examples of unprotected \texorpdfstring{$\AdS_3 \times \Sphere^3 \times \Torus^4$}{AdS3 x S3 x T4} states}
\label{sec:unprot-states}

We illustrate the above ABA construction by presenting a few examples of unprotected $\AdS_3 \times \Sphere^3 \times \Torus^4$ states made up of massless magnon excitations. To begin with we need to study the solutions to the auxiliary Bethe equation~\eqref{eq:BE-aux}. Since physical states satisfy the level-matching constraint\footnote{The square-root branch cuts in~\eqref{eq:BE-aux} are chosen so that the level-matching constraint also implies $\prod_{j=1}^{N_0} \sqrt{x_j^+/x_j^-} = 1$.}
\begin{equation}
  1 = \prod_{j=1}^{N_0} \frac{x_j^+}{x_j^-} \,,
\end{equation}
we can write~\eqref{eq:BE-aux} as
\begin{equation}
  1 = \prod_{j=1}^{N_0} \frac{y_{I,k} - x_j^-}{y_{I,k} - x_j^+} \,.
\end{equation}
For level-matched states, this equation has two solutions independent of the $x^\pm_j$: $y_{I,k} \to \infty$ or $y_{I,k} = 0$. 
For these two solutions the action of the creation operators $\aba{B}^I(y)$ on the state is proportional to that of a supercharge, for example
\begin{equation}\label{eq:B-asympt-sucharge}
\begin{aligned}
  \aba{B}^1(y\gg 1) &= -\frac{2i}{h} \frac{1}{y}\,\gen{Q}^1_{\sL}+\order\Bigl(\frac{1}{y^2}\Bigr) \,, \qquad &
  \aba{B}^1(y\ll 1) &= +\frac{2y}{h} \, \gen{S}^1_{\sR}+\order(y^2)  \,,
  \\
  \aba{B}^3(y\gg 1) &= +\frac{2i}{h} \frac{1}{y}\,\gen{Q}^2_{\sL}+\order\Bigl(\frac{1}{y^2}\Bigr) \,,\qquad &
  \aba{B}^3(y\ll 1)&= -\frac{2y}{h} \, \gen{S}^2_{\sR}+\order(y^2)  \,.
\end{aligned}
\end{equation}
Hence, any state with an auxiliary root $y_{I,k}$ at $0$ or $\infty$ is a descendent. In addition to these two special solutions, the auxiliary Bethe equation has $N_0-2$ additional solutions, which give rise to highest-weight states. Below we consider some simple examples. We have also included a Mathematica notebook which generates complete expressions for wavefunctions of states with up to four magnons and their descendants in the anciliary material submitted with this paper.

\paragraph{Example 1: $N_0=2$. }

We start with a reference state with $N_0 = 2$, for which level-matching means that the two excitations have momenta $p_2=-p_1$
\begin{equation}\label{eq:two-mag-hw}
 \ket{\chi_{p_1} \chi_{p_2}}\,.
\end{equation}
For $N_0=2$ the only solutions to the auxiliary Bethe equations are located at $0$ and $\infty$. The $\aba{B}^I(y_{I,k})$ are now proportional to raising supercharges of $\alg{psu}(1|1)^4_{\ce}$ and acting with them on the above eigenstate generates a long, 16-dimensional  $\alg{psu}(1|1)^4_{\ce}$ multiplet with $N_I=0,1,2$. This matches the representation theory expectation, since each excitation transforms in a short four-dimensional representation and the tensor product of two such representations generically gives a single long 16-dimensional  representation.

For example, acting with a single $\aba{B}^1$ operator on the highest-weight state we find
\begin{equation}\label{eq:n-is-2-single-B-generic}
  \aba{B}^1(y) \ket{\chi_{p_1} \chi_{p_2}}
  \propto
  \frac{\eta_1}{y - x_1^+} \ket{ T^1_{p_1} \chi_{p_2}}
  - \frac{\eta_2}{y - x_2^+} \frac{y - x_1^-}{y - x_1^+} \sqrt{\frac{x_1^+}{x_1^-}} \ket{ \chi_{p_1} T^1_{p_2}} \,.
\end{equation}
At large $y$ this reduces to 
\begin{equation}\label{eq:n-is-2-desc}
  \aba{B}^1(\infty) \ket{\chi_{p_1} \chi_{p_2}}
  \propto
  \frac{\eta_1}{y} \ket{ T^1_{p_1} \chi_{p_2}}
  - \frac{\eta_2}{y} \sqrt{\frac{x_1^+}{x_1^-}} \ket{ \chi_{p_1} T^1_{p_2}} \,,
\end{equation}
and for small $y$ we get
\begin{equation}
  \aba{B}^1(0) \ket{\chi_{p_1} \chi_{p_2}}
  \propto
  -\frac{\eta_1}{x_1^+} \ket{ T^1_{p_1} \chi_{p_2}}
  +\frac{\eta_2}{x_2^+} \sqrt{\frac{x_1^-}{x_1^+}} \ket{ \chi_{p_1} T^1_{p_2}} \,.
\end{equation}
Up to overall normalizations, these two expressions agree with the action on a level-matched $N_0=2$ state of the supercharges $\gen{Q}_{\sL}^1$ and $\gen{S}_{\sR}^1$, respectively. This can be verified by using the coproduct~\eqref{eq:coprod} together with the explicit expressions for the supersymmetry generators given in equation~\eqref{superchargerepsu114}.

As another example, consider acting with one $\aba{B}^1$ operator and one $\aba{B}^3$ operator to find $N_1 = N_3 = 1$  states
\begin{equation}
  \begin{aligned}
    \aba{B}^3(\infty)\aba{B}^1(\infty)\ket{\chi\chi}\propto & \ket{\chi \tilde{\chi}}+f_p^2\ket{\tilde{\chi}\chi}-f_p\ket{T^1T^2}+f_p\ket{T^2T^1} ,\\
    \aba{B}^3(0)\aba{B}^1(\infty)\ket{\chi\chi}\propto & \ket{\chi \tilde{\chi}}-\ket{\tilde{\chi}\chi}-f_p\ket{T^1T^2}-\frac{1}{f_p}\ket{T^2T^1} ,\\
    \aba{B}^3(\infty)\aba{B}^1(0)\ket{\chi\chi}\propto & \ket{\chi \tilde{\chi}}-\ket{\tilde{\chi}\chi}+\frac{1}{f_p}\ket{T^1T^2}+f_p\ket{T^2T^1} ,\\
    \aba{B}^3(0)\aba{B}^1(0)\ket{\chi\chi}\propto & \ket{\chi \tilde{\chi}}+\frac{1}{f_p^2}\ket{\tilde{\chi}\chi}+\frac{1}{f_p}\ket{T^1T^2}-\frac{1}{f_p}\ket{T^2T^1} ,
  \end{aligned}
\end{equation}
where
\begin{equation}
  f_p=\sqrt{x_{p}^-x_{p}^+}
\end{equation}
and we have suppressed the momenta $p_i$  of the two excitations for brevity. As expected for descendants, these states can also be obtained by acting with supercharges\footnote{We label the supercharges of the two copies of $\alg{psu}(1|1)^2_{\ce}$ as $1$ and $2$, but the corresponding auxiliary roots as $1$ and $3$. The latter choice is motivated by the Dynkin diagram of $\alg{psu}(1,1|2)$, which underlies the structure of our Bethe equations as discussed in Figure 7 of~\cite{Borsato:2016xns}.}
\begin{equation}
  \begin{aligned}
    \gen{Q}_{\sL}^2 \gen{Q}_{\sL}^1 \ket{\chi\chi} &\propto 
   \aba{B}^3(\infty)\aba{B}^1(\infty)\ket{\chi\chi} \,, \qquad &
    \gen{S}_{\sR}^2 \gen{Q}_{\sL}^1 \ket{\chi\chi} &\propto
    \aba{B}^3(0)\aba{B}^1(\infty)\ket{\chi\chi} \,, \\
    \gen{Q}_{\sL}^2 \gen{S}_{\sR}^1 \ket{\chi\chi} &\propto \aba{B}^3(\infty)\aba{B}^1(0)\ket{\chi\chi} \,, \qquad &
    \gen{S}_{\sR}^2 \gen{S}_{\sR}^1 \ket{\chi\chi} &\propto \aba{B}^3(0)\aba{B}^1(0)\ket{\chi\chi}\,.
  \end{aligned}
\end{equation}

\paragraph{Example 2: $N_0=3$. }

Next,  consider three-magnon excitations. In addition to the solutions $y=0$ and $y=\infty$, the auxiliary Bethe equation has the additional solution
\begin{equation}\label{eq:N0-3-aux-root}
  y_* = \frac{\sum_{i<j} x_i^+ x_j^+ - \sum_{i<j} x_i^- x_j^-}{\sum_{i} x_i^+ - \sum_{i} x_i^-} \,.
\end{equation}
In the pure RR case this reduces to $y_* = 1$. There are now four highest weight states
\begin{equation}\label{N03hws}
  \ket{\chi_{p_1} \chi_{p_2} \chi_{p_3}} \,, \qquad
  \aba{B}^I(y_*) \ket{\chi_{p_1} \chi_{p_2} \chi_{p_3}} \,, \qquad
  \aba{B}^1(y_*) \aba{B}^3(y_*) \ket{\chi_{p_1} \chi_{p_2} \chi_{p_3}} \,,
\end{equation}
where $p_1+p_2+p_3=0$. Explicitly, we have for example
\begin{equation}\label{eq:state-210-general}
  \begin{aligned}
    \aba{B}^1(y_*) \ket{\chi_{p_1} \chi_{p_2} \chi_{p_3}} 
    &\propto 
    \frac{\eta_1}{y_* - x_1^+} \ket{ T^1_{p_1} \chi_{p_2} \chi_{p_3}}
    - \frac{\eta_2}{y_* - x_2^+} \frac{y_* - x_1^-}{y_* - x_1^+} \sqrt{\frac{x_1^+}{x_1^-}} 
    \ket{ \chi_{p_1} T^1_{p_2} \chi_{p_3}}
    \\ &\qquad
    + \frac{\eta_3}{y_* - x_3^+} \frac{y_* - x_1^-}{y_* - x_1^+} \frac{y_* - x_2^-}{y_* - x_2^+} \sqrt{\frac{x_1^+}{x_1^-} \frac{x_2^+}{x_2^-}} \ket{ \chi_{p_1} \chi_{p_2} T^1_{p_3} } \,.
\end{aligned}
\end{equation}
The remaining highest-weight state is obtained by acting with two $\aba{B}^I(y_*)$ operators and is given explicitly in equation~\eqref{N11N31hwsbasis} in Appendix~\ref{N03N11N31hws}.

As in the $N_0=2$ case, we can fill out the full representations by adding additional roots at $0$ and $\infty$, where the action of $\aba{B}^I(y)$ reduces to that of a supercharge. For example, for $y=\infty$
\begin{equation}\label{eq:n0-is-3-descendant}
  \aba{B}^1( \infty) \ket{\chi_{p_1} \chi_{p_2} \chi_{p_3}}
  \propto 
  \frac{\eta_1}{y} \ket{ T^1_{p_1} \chi_{p_2} \chi_{p_3}}
  - \frac{\eta_2}{y} \sqrt{\frac{x_1^+}{x_1^-}}
  \ket{ \chi_{p_1} T^1_{p_2} \chi_{p_3}}
  + \frac{\eta_3}{y} \sqrt{\frac{x_1^+}{x_1^-} \frac{x_2^+}{x_2^-}} \ket{ \chi_{p_1} \chi_{p_2} T^1_{p_3} } \,,
\end{equation}
and we can check again that this state is a descendant
\begin{equation}
\aba{B}^1( \infty) \ket{\chi_{p_1} \chi_{p_2} \chi_{p_3}}
\propto \gen{Q}_{\sL}^1  \ket{\chi_{p_1} \chi_{p_2} \chi_{p_3}}\,.
\end{equation}
As before, this matches the representation theory of $\alg{psu}(1|1)^4_{\ce}$, since the product of three short representations can be generically decomposed into a sum of four long representations,  $(2|2)^{\otimes 3}=(8|8)^{\oplus 4}$.

\section{Protected states from Bethe ansatz wave functions}\label{T4section}

In this section we find the wave-functions of protected states in $\AdS_3 \times \Sphere^3 \times \Torus^4$ using the ABA constructed in the previous section.  A generic unprotected ABA state of the form discussed in the previous section is built from $N_0$ momentum-carrying roots $p_k$ and $N_I$ auxiliary roots $y_{I,j}$
\begin{equation}
\ket{\vec{p};\vec{y_1};\vec{y_3}}\,,
\end{equation}
as introduced in equation~\eqref{ABAsu114}. Protected states do not receive corrections to their energies and since the dispersion relation~\eqref{eq:disp-rel} depends on the magnon momentum $p_k$, protected states come from zero-momentum magnons only. As we  show below, the auxiliary roots $y_{I,j}$ also take special values for protected states $y_{I,j}=s_{\sPM}$ with $s_{\sPM}$ defined in equation~\eqref{splusminusdef}. In order not to over-load the notation and to distinguish them from unprotected states,  we will label protected states by the number of momentum-carrying and auxiliary roots
\begin{equation}\label{eq:prot-state-not}
\ket{N_0,N_1,N_3}\equiv\ket{\vec{p}=\vec{0};\vec{y_1}=\vec{s}_\pm;\vec{y_3}=\vec{s}_\pm}\,.
\end{equation}
Since the charges of these states follow from equation~\eqref{modevsglobalT4}, we can equivalently write these states in the notation of equation~\eqref{eq:sugra-mplet-notation}
\begin{equation}\label{eq:prot-ket-de-boer}
\ket{N_0,N_1,N_3}\equiv\left(L+N_1+N_3-N_0+1,L+1\right)_{\sS}
\,.
\end{equation}
On the right-hand side above, we have re-introduced the $L$-dependence of the BMN vacuum $\ket{0,0,0}$ that is suppressed in equations like~\eqref{eq:prot-state-not} for compactness.

\subsection{Fermionic zero modes}

In order to find the protected states, we send magnon momenta  to zero. Since the massless dispersion relation~\eqref{eq:disp-rel} has a cusp at $p=0$ we get potentially different results if we send $p \to 0$ from above or below. For $p=0$, the Zhukovski variables satisfy $x^+_p=x^-_p$ and we will denote by $s_{\sP}$ or $s_{\sM}$ their value as $p$ goes to zero from above or below
\begin{equation}\label{splusminusdef}
  s_{\sP} \equiv \lim_{p \to 0^+} x_p^{\pm} = \frac{\k + \sqrt{\k^2 + h^2}}{h} \,, \qquad
  s_{\sM} \equiv \lim_{p \to 0^-} x_p^{\pm} = \frac{\k - \sqrt{\k^2 + h^2}}{h} = - \frac{1}{s_{\sP}} \,,
\end{equation}
which become $\pm 1$ for $\k=0$. 

The simplest zero-momentum states have no auxiliary excitations and setting $x_\pm=s_{\sP}$ or $x^\pm=s_{\sM}$ leads to the same state, hence denoting the corresponding excitation by $\chi_{p=0}$ is unambiguous. For example, a single excitation ($N_0 = 1$) with zero momentum\footnote{Because of the level-matching constraint $\sum_i p_i = 0$, for $N_0 = 1$  we have to set $p_1=0$.}  written in the notation of equation~\eqref{eq:prot-state-not} is
\begin{equation}\label{eq:n0-is-one-prot}
\ket{1,0,0}^{\dot{a}}=
  \ket{\chi_0^{\dot{a}}} \,,
\end{equation}
where we reintroduced the $\alg{su}(2)_{\circ}$ index of $\chi$, as discussed in footnote~\ref{ft:su2-circ}. Hence, the $N_0 = 1$ reference state gives an $\alg{su}(2)_{\circ}$ doublet of protected states. Similarly, for $N_0 = 2$ we have
\begin{equation}\label{eq:n0-is-two-inter}
\ket{2,0,0}^{\dot{a}\dot{b}}=  \ket{\chi^{\dot{a}}_0 \chi^{\dot{b}}_0} \,.
\end{equation}
This  is not yet a full physical state, since we still have to sum over all permutations of the excitations as described in equation~\eqref{eq:total-2mag-state}. At zero momentum, the S matrix reduces to a graded permutation, so the full state is given by
\begin{equation}\label{eq:n0-is-two-prot}
\epsilon_{\dot{a}\dot{b}}\ket{2,0,0}^{\dot{a}\dot{b}}=   \ket{\chi^{\dot{a}}_0 \chi^{\dot{b}}_0} - \ket{\chi^{\dot{b}}_0 \chi^{\dot{a}}_0} \,.
\end{equation}
Hence, the $N_0=2$ reference state give rise to a single protected state, which is an $\alg{su}(2)_{\circ}$ singlet.

The remaining protected states carry auxiliary roots. But where should these roots sit? If we take the auxiliary Bethe equation~\eqref{eq:BE-aux} and send all momenta to zero, we find it is trivially satisfied for any $y$. Naively then we might conclude that acting with $\aba{B}^I(y)$ for any $y$  gives a physical protected state. However, this cannot be correct, since at  $y=0$, $\infty$ the $\aba{B}^I(y)$ reduce to $\alg{psu}(1|1)^4_{\ce}$ supercharges (see equation~\eqref{eq:B-asympt-sucharge}), which by definition annihilate all protected states. As we show below, there are natural values for $y$ to take for protected states.

To determine which value the auxiliary roots need to take, consider the $N_0 = 3$ states discussed in Section~\ref{sec:unprot-states} in the limit $p_1\rightarrow 0$, for which $x_1^{\pm}$ goes to either $s_{\sP}$ or $s_{\sM}$ (see equation~\eqref{splusminusdef}). In this limit the auxiliary root  solution~\eqref{eq:N0-3-aux-root} becomes
\begin{equation}
  y_* = s_{\sPM} + \frac{x_2^+ x_3^+ - x_2^- x_3^-}{x_2^+ + x_3^+ - x_2^- - x_3^-} + \order(p_1)\,.
\end{equation}
Since the level-matching condition reduces to $p_2+p_3=0$, for which $x_2^+ x_3^+ - x_2^- x_3^-=0$, the auxiliary root for $p_1=0$ is
\begin{equation}\label{eq:special-val-aux-root}
 y_* = s_{\sPM}\,.
 \end{equation}
 In other words, in this limit the auxiliary root takes the \textit{same} value as the momentum-carrying root 
 \begin{equation}\label{aux-root-prot-state}
 y_{I,1}=x_1^+=x_1^-=s_{\sPM}\,.
 \end{equation}

 For example, consider the  highest-weight state in equation~\eqref{eq:state-210-general}.
Inserting $y_*=s_{\sPM}$ and sending $p_1\rightarrow 0$, we see that the coefficient of the first term on the right-hand side of~\eqref{eq:state-210-general} diverges, since $\eta_{p_1} = \order(\sqrt{p_1})$, while the remaining two terms stay finite. For a normalized state  this means that we can neglect the latter terms in the limit, leading to 
\begin{equation}
\lim_{p_1\rightarrow 0}\ket{p_1,p_2,p_3;y_{1,1}=y_*}=\ket{0,p_2,p_3;s_{\sPM}}\propto \ket{T_0^1 \chi_{p_2} \chi_{-p_2}}\,.
\end{equation}
This is a highest-weight two-magnon state, similar to the state in equation~\eqref{eq:two-mag-hw}, but with an additional $T_0^1$ zero mode. While such \textit{bosonic} zero modes will not play a role in the protected spectrum, this example illustrates an important feature of the special value of the auxiliary root $y=s_{\sPM}$: the operator $\aba{B}(s_{\sPM})$ acts \textit{only} on the corresponding zero momentum site changing the highest weight state $\chi_{p=0}$ to a $\rho_{\alg{psu}(1|1)^4}$ descendant, in this case $T_{p=0}^1$.

Similarly, if we considered the  highest-weight state~\eqref{N11N31hwsbasis} in the $p_1\rightarrow 0$ limit, we find
\begin{equation}\label{eq:zero-mom-limit-hw}
\lim_{p_1\rightarrow 0}\ket{p_1,p_2,p_3;y_{1,1}=y_*;y_{3,1}=y_*}=\ket{0,p_2,-p_2;s_{\sPM};s_{\sPM}}\propto \ket{{\tilde\chi}_0 \chi_{p_2} \chi_{-p_2}}\,.
\end{equation}
This too is a highest-weight two magnon state, similar to the state in equation~\eqref{eq:two-mag-hw}, now with an additional ${\tilde\chi}_0$ zero mode. As we discuss in the next subsection,  sending $p_2\rightarrow 0$ will lead to further protected states. 

The above examples illustrate how our ABA takes into account the further ground-states which exist in $\AdS_3$ integrable backgrounds in addition to the BMN vacuum  on which the generic state~\eqref{ABAsu114} is constructed. For such a state, sending one of the momenta $p_k$ to zero and setting  one  auxiliary root $y_{I,j}$ to $s_{\sPM}$ leads to a state with $N_0-1$ magnons and $N_I-1$ auxiliary roots in the presence of a massless bosonic zero-mode, while setting $y_{1,j}=y_{3,j}=s_{\sPM}$ gives a state with $N_0-1$ magnons as well as $N_1-1$ and $N_3-1$ auxiliary roots.

On the other hand, states whose auxiliary roots are not $s_{\sPM}$ in a zero-momentum limit, become states with $N_0-1$ magnons, $N_I$ auxiliary roots and a $\chi_0$ zero-mode, which is highest-weight in $\rho_{\alg{psu}(1|1)^4}(p_k=0)$. For example, consider the $N_0=3$ descendant state in equation~\eqref{eq:n0-is-3-descendant}. There, the first term is suppressed in the $p_1 \to 0$ limit by a factor $\sqrt{p_1}$, giving 
\begin{equation}
\lim_{p_1\rightarrow 0}\ket{p_1,p_2,p_3;y_{1,1}=\infty}=\ket{0,p_2,p_3;\infty}\propto 
-\eta_2\ket{\chi_0 T^1_{p_2} \chi_{p_3}}
+\eta_3\sqrt{\frac{x_2^+}{x_2^-}}\ket{\chi_0 \chi_{p_2}T^1_{p_3}}
\,,
\end{equation}
with $p_3=-p_2$ due to level-matching. This state is a descendant of the highest-weight state $\ket{\chi_0 \chi_{p_2}\chi_{-p_2}}$ and in this way is similar to the $N_0=2$ descendant in equation~\eqref{eq:n-is-2-desc}, with an additional $\chi_0$ zero mode inserted. In the $p_2\to 0$ limit, this state goes to zero as expected for a $\alg{psu}(1|1)^4_{\ce}$ descendant. This example illustrates the fact that, in the zero-momentum limit, states whose auxiliary roots do not become $s_{\sPM}$ will not give rise to protected states. This example illustrates a second important feature of the massless ABA: in the limit where one momentum goes to zero, the operator $\aba{B}(y_{I,j})$ for $y_{I,j}\neq s_{\sPM}$ acts on all sites \textit{except} the  zero momentum site.

From the above discussion, we conclude that protected states involving ${\tilde\chi}_0$ zero modes have auxiliary roots at $y = s_{\sPM}$ as $p \to 0^\pm$. For the $N_0=2$ state~\eqref{eq:n0-is-two-inter}, we can add at most two auxiliary roots of each $I$-type\footnote{Since we are only interested in fermionic zero modes we always turn on the auxiliary roots pairwise, with one root of type 1 and one root of type 3 sitting at the same point.}
\begin{equation}\label{eq:prot-two-chi-tildes}
  \ket{2,2,2}^{\dot{a}\dot{b}} = \aba{B}^1(s_{\sP}) \aba{B}^3(s_{\sP}) \aba{B}^1(s_{\sM}) \aba{B}^3(s_{\sM})  \ket{\chi^{\dot{a}}_0 \chi^{\dot{b}}_0} \sim \ket{ \tilde{\chi}^{\dot{a}}_0 \tilde{\chi}^{\dot{b}}_0} ,
\end{equation}
which leads to a single protected state 
\begin{equation}
\epsilon_{\dot{a}\dot{b}} \ket{2,2,2}^{\dot{a}\dot{b}}=\epsilon_{\dot{a}\dot{b}}  \ket{ \tilde{\chi}^{\dot{a}}_0 \tilde{\chi}^{\dot{b}}_0}
\end{equation}
which is a singlet under $\alg{su}(2)_{\circ}$ once we sum over all permutations of the excitations as  in equation~\eqref{eq:total-2mag-state}. In equation~\eqref{eq:prot-two-chi-tildes}, we have chosen a particular ordering of $s_{\sP}$ and $s_{\sM}$ in the arguments of the $\aba{B}$ operators as well as implicitly setting $x^\pm_1=s_{\sP}$ and $x^\pm_2=s_{\sM}$. One can check that different choices of $s_{\sPM}$ lead to the same protected states as long as this is done consistently with equation~\eqref{aux-root-prot-state}.

More interesting is the case where we turn on one auxiliary root of each type. We now have two possibilities\footnote{Other combinations of $\aba{B}$ operators, such as $\aba{B}^1(s_{\sP}) \aba{B}^3(s_{\sM})$ would lead to states with two zero-momentum bosons instead of fermions.}
\begin{equation}
 \aba{B}^1(s_{\sP}) \aba{B}^3(s_{\sP}) \ket{\chi^{\dot{a}} \chi^{\dot{b}}} \sim \ket{ \tilde{\chi}^{\dot{a}} \chi^{\dot{b}}} \qquad\text{or}\qquad 
 \aba{B}^1(s_{\sM}) \aba{B}^3(s_{\sM}) \ket{\chi^{\dot{a}} \chi^{\dot{b}}} \sim \ket{ \chi^{\dot{a}} \tilde{\chi}^{\dot{b}}} .
\end{equation}
These states are described by two distinct sets of Bethe roots, and at first sight we would expect them to represent different states. However,  summing over all permutations of the excitations as in equation~\eqref{eq:total-2mag-state} gives
\begin{equation}
  \ket{ \tilde{\chi}_0^{\dot{a}} \chi_0^{\dot{b}}} - \ket{ \chi_0^{\dot{b}} \tilde{\chi}_0^{\dot{a}}}  \qquad\text{or}\qquad 
  \ket{ \chi_0^{\dot{a}} \tilde{\chi}_0^{\dot{b}}} - \ket{ \tilde{\chi}_0^{\dot{b}} \chi_0^{\dot{a}}} ,
\end{equation}
which are the same states. The protected states at this level can thus be decomposed into a triplet and a singlet of $\alg{su}(2)_{\circ}$. Hence, we find that \emph{two different solutions to the Bethe equations can lead to the same physical protected state}.

We end this subsection by noting that when performing explicit computations of how the $\aba{B}(s_{\sPM})$ operators act on states of the form~\eqref{eq:n0-mag-pseud-state} it is important to take an \textit{ordered} zero-momentum limit $0\leftarrow |p_1|<|p_2|<\dots$. This is because, as we discussed below equation~\eqref{eq:zero-mom-limit-hw}, one needs to re-normalize the eigenstates as each momentum reaches zero. One can check that in the end different orderings lead to the same protected eigenstates, but at intermediate stages of the calculation care must be taken with expressions involving multiple $\aba{B}$ operators. 

In the next subsection we  summarize the complete set of protected states obtained by picking an ordering of the  $\aba{B}$ operators, as well as making particular choices for the each zero-momentum Zhukovsky variable as discussed in equation~\eqref{splusminusdef} in a way that is consistent with equation~\eqref{aux-root-prot-state}. We have checked that all other allowed choices lead to the same physical protected eigenstates just as the $N_0=2$ cases described above.

\subsection{Protected states}

In this sub-section, we write down the complete closed-string protected spectrum for $\AdS_3\times \Sphere^3\times \Torus^4$, using the ABA procedure described in detail above. As discussed there, the Bethe equations' solutions corresponding to these states are not unique, and for each protected state we have the freedom to choose a useful way of taking the zero momentum limit. All such choices lead to the same physical protected states. Below we make one such choice and list the states ordered by the number of zero modes they contain. Furthermore, we do not explicitly write down the sum over permutations of the zero modes, as described in equation~\eqref{eq:total-2mag-state}, for simplicity listing a single representative for each state.\footnote{For a state with only zero modes, the sum over permutations just gives the obvious fermionic minus signs.}

\subsubsection{\texorpdfstring{$N_0=0$}{N0 = 0}}

Without any zero modes we have just the family of BMN ground states
\begin{equation}
  \ket{0,0,0} ,
\end{equation}
labelled by $L$ with $2D_{\sL}=2D_{\sL}=2J_{\sL}=2J_{\sR}=L$, which are $\algSU(2)_{\circ}$ singlets.

\subsubsection{\texorpdfstring{$N_0=1$}{N0 = 1}}

The protected states with a single $\chi^{\dot{a}}$ zero-mode were introduced in~\eqref{eq:n0-is-one-prot}
\begin{equation}\label{state100}
  \ket{1,0,0}^{\dot{a}} = \ket{\chi_0^{\dot{a}}}\,,
\end{equation}
while those with a single $\tilde{\chi}^{\dot{a}}$ zero-mode have two $\aba{B}$ operators
\begin{equation}\label{state111}
  \ket{1,1,1}^{\dot{a}} = \aba{B}^1(s_{\sP}) \aba{B}^3(s_{\sP}) \ket{1,0,0} \propto \ket{\tilde{\chi}_0^{\dot{a}}} \,.
\end{equation}
Both types of states are $\algSU(2)_{\circ}$ doublets.

\subsubsection{\texorpdfstring{$N_0=2$}{N0 = 2}}

Here we start with the reference state introduced in~\eqref{eq:n0-is-two-prot}
\begin{equation}
\epsilon_{\dot{a}\dot{b}}  
\ket{2,0,0}^{\dot{a}\dot{b}} = \epsilon_{\dot{a}\dot{b}}  
\ket{ \chi_{0^+}^{\dot{a}} \chi_{0^-}^{\dot{b}} } ,
\end{equation}
which is a $\algSU(2)_{\circ}$-singlet. Adding one set of $\aba{B}$ operators we have\footnote{In obtaining this state, it is important to first act with the $\aba{B}$ operators and then sum over permutations, as described in equation~\eqref{eq:total-2mag-state}.}
\begin{equation}
  \ket{2,1,1}^{\dot{a}\dot{b}} =  \aba{B}^1(s_{\sP}) \aba{B}^3(s_{\sP}) \ket{2,0,0}^{\dot{a}\dot{b}} \propto   \ket{ \tilde{\chi}_{0^+}^{\dot{a}} \chi_{0^-}^{\dot{b}} } \,,
\end{equation}
which is a triplet and a singlet under $\algSU(2)_{\circ}$. With two sets of $\aba{B}$ operators we find
\begin{equation}
\epsilon_{\dot{a}\dot{b}}  
  \ket{2,2,2}^{\dot{a}\dot{b}}  
 =\epsilon_{\dot{a}\dot{b}}   \aba{B}^1(s_{\sP}) \aba{B}^3(s_{\sP}) \aba{B}^1(s_{\sM}) \aba{B}^3(s_{\sM}) 
 \ket{2,0,0}^{\dot{a}\dot{b}}  
  \propto \epsilon_{\dot{a}\dot{b}}  
\ket{ \tilde{\chi}_{0^+}^{\dot{a}} \tilde{\chi}_{0^-}^{\dot{b}} }^{\dot{a}\dot{b}}  
\,,
\end{equation}
which is  a $\algSU(2)_{\circ}$ singlet.

\subsubsection{\texorpdfstring{$N_0=3$}{N0 = 3}}

With three fermions the reference state would take the form
\begin{equation}
  \ket{ \chi_{0^+}^{\dot{a}} \chi_{0^-}^{\dot{b}} \chi_{0^-}^{\dot{c}} } .
\end{equation}
However, since the fermions are doublets under $\algSU(2)_{\circ}$, fermion statistics will kill the corresponding full physical state. To get a non-vanishing state we need to include at least one set of $\aba{B}$  operators
\begin{equation}
  \ket{3,1,1}^{\dot{a}} = \aba{B}^1(s_{\sP}) \aba{B}^3(s_{\sP}) \epsilon_{\dot{b}\dot{c}} \ket{ \chi_{0^+}^{\dot{a}} \chi_{0^-}^{\dot{b}} \chi_{0^-}^{\dot{c}} }
  \propto \epsilon_{\dot{b}\dot{c}} \ket{ \tilde{\chi}_{0^+}^{\dot{a}} \chi_{0^-}^{\dot{b}} \chi_{0^-}^{\dot{c}} } ,
\end{equation}
We can also include two sets of $\aba{B}$  operators to get
\begin{equation}\label{eq:ket-322}
  \ket{3,2,2}^{\dot{a}} = \aba{B}^1(s_{\sP}) \aba{B}^3(s_{\sP}) \aba{B}^1(s_{\sM}) \aba{B}^3(s_{\sM}) \epsilon_{\dot{b}\dot{c}} \ket{ \chi_{0^+}^{\dot{a}} \chi_{0^+}^{\dot{b}} \chi_{0^-}^{\dot{c}} }
  \propto \epsilon_{\dot{b}\dot{c}} \ket{ \tilde{\chi}_{0^+}^{\dot{a}} \tilde{\chi}_{0^-}^{\dot{b}} \chi_{0^-}^{\dot{c}} } .
\end{equation}
Both types of states are $\algSU(2)_{\circ}$  doublets. A state with more than two sets of $\aba{B}$  operators would again vanish because of fermion statistics.

\subsubsection{\texorpdfstring{$N_0=4$}{N0 = 4}}

The four fermion zero momentum reference state
\begin{equation}
  \ket{ \chi_{0^+}^{\dot{a}} \chi_{0^+}^{\dot{b}} \chi_{0^-}^{\dot{c}} \chi_{0^-}^{\dot{d}} }\,,
\end{equation}
for which the full physical state vanishes due to Fermi statistics. To get a physical state we now need to include two sets of $\aba{B}$ operators
\begin{equation}
  \begin{aligned}
    \ket{4,2,2}
    &= \aba{B}^1(s_{\sP}) \aba{B}^3(s_{\sP}) \aba{B}^1(s_{\sM}) \aba{B}^3(s_{\sM}) \epsilon_{\dot{a}\dot{b}} \epsilon_{\dot{c}\dot{d}}  \ket{ \chi_{0^+}^{\dot{a}} \chi_{0^+}^{\dot{b}} \chi_{0^-}^{\dot{c}} \chi_{0^-}^{\dot{d}} }
    \\
    &\propto \epsilon_{\dot{a}\dot{b}} \epsilon_{\dot{c}\dot{d}} \ket{ \chi_{0^+}^{\dot{a}} \tilde{\chi}_{0^+}^{\dot{b}} \tilde{\chi}_{0^-}^{\dot{c}} \chi_{0^-}^{\dot{d}} } \,,
\end{aligned}
\end{equation}
which is a $\algSU(2)_{\circ}$ singlet.

\subsubsection{Summary of protected states}

The protected states found in this section are summarised below, where  superscripts  indicate the $\algSU(2)_{\circ}$ representations
\begin{equation}\label{eq:summary-prot-states-t4}
  \begin{gathered}
    \ket{0,0,0}^{\mathrlap{\rep{1}}}
    \\[2pt]
    \ket{1,0,0}^{\mathrlap{\rep{2}}}
    \hspace{1cm}
    \ket{1,1,1}^{\mathrlap{\rep{2}}}
    \\[2pt]
    \ket{2,0,0}^{\mathrlap{\rep{1}}}
    \hspace{1cm}
    \ket{2,1,1}^{\mathrlap{\rep{1}\oplus\rep{3}}}
    \hspace{1cm}
    \ket{2,2,2}^{\mathrlap{\rep{1}}}
    \\[2pt]
    \ket{3,1,1}^{\mathrlap{\rep{2}}}
    \hspace{1cm}
    \ket{3,2,2}^{\mathrlap{\rep{2}}}
    \\[2pt]
    \ket{4,2,2}^{\mathrlap{\rep{1}}}
  \end{gathered}
\end{equation}
These states match the Hodge diamond of the seed $\Torus^4$ theory~\eqref{eq:seed-t4} and, since they depend additionally on $L$ through the  BMN vacuum $\ket{0,0,0}$ (see equation~\eqref{eq:prot-ket-de-boer}), we match the  expected protected spectrum in~\eqref{eq:t4-prot-spec-de-boer-notation}.\footnote{Recall that, as we have shown in this sub-section, $(N_0,N_1+N_3)$ can only take the following values $(0,0),(1,0),(1,2),(2,0),(2,2),(2,4),(3,2),(3,4),(4,4)$. Since $L\in\mathbb{N}^+$, we have additionally shifted $L$ by suitable discrete amounts when matching the two expressions.}

\section{Protected states in \texorpdfstring{$\AdS_3\times \Sphere^3\times \Kthree$}{AdS3 x S3 x K3} orbifolds}
\label{sec:prot-stat-orb}

In this section, we discuss the ABA spectrum of strings in $\AdS_3\times \Sphere^3\times \Torus^4/\Integers_n$ orbifolds, with  $n=2,3,4,6$. As reviewed in Section~\ref{sec:prot-spec}, $\Integers_n$ acts only on  $\alg{su}(2)_{\circ}$.  Since the massive sector is not charged under the $\alg{su}(2)_{\circ}$ symmetry, there is no impact of the action of the orbifold group on the massive ABA states, nor on the massive Bethe equations~\cite{Borsato:2013qpa}. In particular, the massive excitation $Z$ which generates the BMN-like vacuum $\ket{Z^L}$, is invariant under the orbifold action. 

On the other hand,  massless modes transform as doublets under $\alg{su}(2)_{\circ}$. Since we are only considering states with zero winding and momentum on $\Torus^4$, the untwisted sector states simply have to be invariant under the projection~\eqref{orbifoldaction}. In particular, protected states listed in equation~\eqref{eq:summary-prot-states-t4} survive the $\Integers_{n>2}$ projections only if they are $\alg{su}(2)_{\circ}$ singlets. The $n>2$ untwisted sector protected spectrum is
\begin{equation}\label{eq:summary-prot-states-zn-untw-n-bigger-2}
  \begin{gathered}
    \ket{0,0,0}
    \\[2pt]
    \mbox{\O}
    \hspace{2.1cm}
    \mbox{\O}
    \\[2pt]
    \ket{2,0,0}
    \hspace{1cm}
    \ket{2,1,1}
    \hspace{1cm}
    \ket{2,2,2}
    \\[2pt]
    \mbox{\O}
    \hspace{2.1cm}
    \mbox{\O}
    \\[2pt]
    \ket{4,2,2}
  \end{gathered}
\end{equation}
Compared to~\eqref{eq:summary-prot-states-t4}, we have dropped the superscript denoting the $\algSU(2)_{\circ}$ representations, since  $\algSU(2)_{\circ}$ is broken by the orbifold and each multiplet above has multiplicity one. For $n=2$, the orbifold acts  as a minus sign for each $\alg{su}(2)_\circ$ doublet index and as a result  all four $\ket{2,1,1}$ multiplets are  $\Integers_2$-invariant. The $n=2$ untwisted sector protected spectrum is
\begin{equation}\label{eq:summary-prot-states-zn-untw-n-is-2}
  \begin{gathered}
    \ket{0,0,0}
    \\[2pt]
    \mbox{\O}
    \hspace{2.1cm}
    \mbox{\O}
    \\[2pt]
    \ket{2,0,0}
    \hspace{1.4cm}
    \ket{2,1,1}^{\oplus 4}
    \hspace{1cm}
    \ket{2,2,2}
    \\[2pt]
    \mbox{\O}
    \hspace{2.1cm}
    \mbox{\O}
    \\[2pt]
    \ket{4,2,2}
  \end{gathered}
\end{equation}

Next consider the twisted sectors. The massive ABA and Bethe equations remain unchanged, since those excitations are $\alg{su}(2)_{\circ}$ singlets. In fact, since the orbifold leaves the $\alg{psu}(1|1)^4_{\ce}$ symmetry invariant, only the massless momentum-carrying Bethe equations change. The twisted-sector boundary conditions are implemented in the Bethe equations by an additional phase
\begin{equation}\label{eq:tw-mom-bethe-eq}
  \left(\frac{x_k^+}{x_k^-}\right)^L=e^{-i\phi_0}\prod_{\substack{j=1\\ j \neq k}}^{N_0} \sqrt{\frac{x_k^-}{x_k^+}\frac{x_j^+}{x_j^-}} \frac{x_k^+-x_j^-}{x_k^--x_j^+} (\sigma_{kj}^{\circ\circ})^2 \prod_{j=1}^{N_1} \sqrt{\frac{x_k^+}{x_k^-}} \frac{x_k^--y_{1,j}}{x_k^+-y_{1,j}}\prod_{j=1}^{N_3} \sqrt{\frac{x_k^+}{x_k^-}} \frac{x_k^--y_{3,j}}{x_k^+-y_{3,j}}\,.
\end{equation}
where $\phi_0$ is 
\begin{equation}\label{twistanglesZn}
    \phi_0=\pm\frac{2\pi}{n}
\,,\end{equation}
with the sign choice determined by the value of the $\alg{su}(2)_\circ$ index of the momentum-carrying massless excitation in question. Because of this phase, in the zero momentum limit, there are no $N_0>0$ solutions to the above  Bethe equation and only the BMN vacuum state $\ket{0,0,0}\,\,$ without any fermionic zero modes is protected in each twisted sector. As a result, the counting of pretected multiplets in the twisted sectors matches the Hodge number counting reviewed in Section~\ref{sec:prot-spec}: there are 16, respectively 18, twisted sector multiplets in the $\Integers_2$, respectively $\Integers_{n>2}$, orbifolds. Combining these with the untwisted sector states above, we
match the Hodge diamond of the seed $\Kthree$ theory~\eqref{eq:seed-k3}. Remembering the additional dependence on $L$ through the  BMN vacuum $\ket{0,0,0}$ (see equation~\eqref{eq:prot-ket-de-boer}), we match the  expected protected spectrum in~\eqref{eq:k3-prot-spec-de-boer-notation}.

\section{Protected states in \texorpdfstring{$\AdS_3\times \Sphere^3\times \Sphere^3\times\Sphere^1$}{AdS3 x S3 x S3 x S1} and its orbifold}
\label{sec:s3-and-orb-prot}

The ABA and Bethe equations for closed strings on $\AdS_3\times \Sphere^3\times \Sphere^3\times\Sphere^1$ are based on a single $\alg{psu}(1|1)^2_{\ce}$ algebra. There are two massless $\rho_{\sL}$ multiplets,  with highest-weight fermions denoted as $\chi_{\sL}$ and $\chi_{\sR}$. There is now a single auxiliary variable $y$, whose Bethe equation is the same as~\eqref{eq:BE-aux}. The two massless momentum-carrying Bethe equations are like equation~\eqref{eq:BE-momentum-carrying}, but with only a single auxiliary-root product. As already shown in~\cite{Baggio:2017kza}, the protected states come from the BMN vacuum with up to two fermionic zero modes inserted
\begin{equation}
\ket{Z^L}\,,\qquad\ket{Z^L\chi_{\sL}}\,,\qquad\ket{Z^L\chi_{\sR}}\,,\qquad\ket{Z^L\chi_{\sL}\chi_{\sR}}\,,
\end{equation}
where, as in the $\Torus^4$ case, the Zhukovski variables can take either of the two values $s_{\sP}$ in equation~\eqref{splusminusdef} to give the same physical state. Unlike the $\Torus^4$ case, here there are no protected multiplets involving auxiliary roots. The spectrum found in this way~\cite{Baggio:2017kza} agrees with the supergravity calculation and WZW point analysed in~\cite{Eberhardt:2017fsi}.

When $\alpha=1/2$ the two $\Sphere^3$s in the geometry have the same radius and one can define a $\Integers_2$ orbifold whose action swaps the two three-spheres and inverts the circle~\cite{Yamaguchi:1999gb}. These authors showed that it is possible to extend the orbifold action in a way that preserves $\superN=(3,3)$ super-conformal symmetry, whose global symmetry is $\alg{osp}(3|2)\subset\alg{osp}(4|2)\equiv \alg{d}(2,1;\alpha=\tfrac{1}{2})$. The protected spectrum of this orbifold was found in~\cite{Eberhardt:2018sce}. The $\alg{psu}(1|1)^2_{\ce}$ algebra used to construct the ABA of the unorbifolded background is contained in the $\alg{osp}(3|2)$ symmetry of the orbifolded background. This suggests that the orbifolded theory could also be integrable, with the same ABA as the parent theory, if suitable integrability-preserving quasi-periodicity conditions for the twisted sectors can be found.  Further, since the $\alg{psu}(1|1)^2_{\ce}$ algebra is invariant under the orbifold action so too must be the $\alg{psu}(1|1)^2_{\ce}$ supercurrents given in equations (2.38)--(2.41) of~\cite{Borsato:2015mma}. The massless bosons are the  $S^1$ boson $w$ and the difference of the great circles on the two $\Sphere^3$s $\psi$. The $\Integers_2$  action on these is 
\begin{equation}
w\longrightarrow -w\,,\qquad\qquad
\psi\longrightarrow -\psi\,,
\end{equation}
and in order to preserve the supercurrents the orbifold must act as a 
\begin{equation}
\chi_{\sL}\longrightarrow -\chi_{\sL}\,,\qquad\qquad
\chi_{\sR}\longrightarrow -\chi_{\sR}
\end{equation}
on the massless fermions. The untwisted protected spectrum of $\AdS_3\times(\Sphere^3\times\Sphere^3\times\Sphere^1)/\Integers_2$
 follows immediately: the states
\begin{equation}
\ket{Z^L\chi_{\sL}}\,,\qquad\ket{Z^L\chi_{\sR}}\,,
\end{equation}
are projected out, while
\begin{equation}
\ket{Z^L}\,,\qquad\ket{Z^L\chi_{\sL}\chi_{\sR}}\,,
\end{equation}
survive the projection. The orbifold also has two twisted sectors, one each for the two $\Sphere^1$ fixed points. Each twisted sector's ground state gives rise to a single protected state for fixed $L$. This is to be be expected in the twisted ABA which does not contain any zero-momentum massless fermions, much like in the case of the twisted sectors in $\Torus^4/\Integers_n$ orbifolds. This protected spectrum agrees with the one found in~\cite{Eberhardt:2018sce} using supergravity and WZW methods.

\section{Conclusions}
\label{sec:conclusions}

In this paper, we have constructed the exact ABA for closed strings on $\AdS_3\times\Sphere^3\times\Torus^4$ and $\AdS_3\times\Sphere^3\times\Kthree$ in its orbifold limits. The ABA is valid for these geometries supported by any combination of NSNS and RR charges as well as any value of non-blow up moduli  since all such theories are integrable~\cite{OhlssonSax:2018hgc}. Because of the relatively low amount of supersymmetry, these theories have multiple families of groundstates in addition to the familiar BMN vacua. We have shown how generic  closed string states can be constructed in the ABA on top of each of these groundstates by inserting fermionic zero-modes. This novel feature, not found in higher-dimensional integrable holographic models, relies on the presence of massless momentum carrying roots, as well as new special points in the auxiliary Bethe variables. 

As we showed in equation~\eqref{aux-root-prot-state}, starting with a Bethe state containing a massless momentum-carrying excitation at zero momentum $x^\pm= s_{\sPM}$, we can create another state by acting with a $\aba{B}$ operator with auxiliary root $y=s_{\sPM}$. This leads to an excited state on top of a different vacuum with other zero-modes inserted. Such a feature of the ABA is unique for $m=0$ momentum carrying excitations, because the zero-momentum limit in equation~\eqref{splusminusdef} gives a finite answer only when $m=0$. This should be contrasted with  adding auxiliary roots at $\pm\infty$ to a generic Bethe state~\cite{Beisert:2005fw}. 
In this more familiar case,  the resulting states are descendents in the same multiplet of the global superconformal algebra, in contrast to the massless states with zero-momentum roots, which are in different multiplets.


Using the ABA we constructed the explicit low-magnon number eigenstates and showed how, in the zero-momentum limit, they give rise to protected  half-BPS states in these backgrounds. The protected spectrum was first found using supergravity Kaluza-Klein reduction in~\cite{deBoer:1998kjm} and we reproduce it with the ABA here. Our analysis proves that such states are  protected to all orders in $\alpha'$ in the planar theory and amounts to a non-renormalization theorem based on integrability. By constructing the explict wavefunctions of  states, we were able to perform the $\Torus/\Integers_n$ orbifolds in a straightforward way, obtaining for the first time the protected spectrum of the $\AdS_3\times\Sphere^3\times\Kthree$ theories from exact integrable methods. Additionally, our wavefunction construction demonstrates explicitly that each protected state can be obtained from a number of distinct solutions of the Bethe equations, as was  anticipated in~\cite{Baggio:2017kza}.

It is not clear to us whether integrability continues to hold away from the orbifold limit of $\Kthree$. While this would not affect the protected spectrum analysis, since blow-up modes are moduli, a generic $\Kthree$ has a complicted metric making the integrability of string theory on it less likely. To test this, one might, for example, compute the magnon dispersion relation  or  generalise the analysis of~\cite{Wulff:2019tzh} in the presence of a few blow-up mode insertions. At the same time, as the analysis in~\cite{OhlssonSax:2018hgc} showed, integrability is valid across the whole moduli space of 
$\AdS_3\times\Sphere^3\times\Torus^4$ theories. We might optimistically hope for the same to be true for $\AdS_3\times\Sphere^3\times\Kthree$.

We have also generalised our analysis of the protected string spectrum to the quarter-BPS protected states for the  $\AdS_3\times \Sphere^3\times \Sphere^3\times\Sphere^1$ background and its $\Integers_2$ orbifold. The protected eigenstates are much simpler since they do not involve any auxiliary roots and for the unorbifolded theory was already presented in~\cite{Baggio:2017kza}.

It would be interesting to use these results to better understand the Thermodynamic Bethe Ansatz (TBA) for these backgrounds, building on the massless TBA of the pure RR theory found in~\cite{Bombardelli:2018jkj}. The role of the auxiliary roots we have presented may  help to clarify how  mixed-mass interactions can be incorporated into the TBA following also the observations in~\cite{Abbott:2020jaa}.

\section*{Acknowledgements}

We would like to thank Juan Miguel Nieto for helpful comments on our manuscript. SM is funded by SMCSE Doctoral Studentship at City University. BS acknowledges funding support from  STFC Consolidated Grants `Theoretical Physics at City University' ST/P000797/1 and ST/T000716/1. BS is grateful for the hospitality of Perimeter Institute where part of this work was carried out. Research at Perimeter Institute is supported in part by the Government of Canada through the Department of Innovation, Science and Economic Development Canada and by the Province of Ontario through the Ministry of Economic Development, Job Creation and Trade. AT is supported by the EPSRC-SFI grant EP/S020888/1 \textit{Solving Spins and Strings}.
\appendix

\section{Derivation of Bethe equations in Zhukovski variables}\label{app:Betheeqproofs}

In this section, we  derive the Bethe equations~\eqref{eq:BE-aux}\,,~\eqref{eq:BE-momentum-carrying}. We  show this for a single copy of $\alg{psu}(1|1)^2_{\ce}$, labelled $I=1$. The proof works the same way for the other copy of $\alg{psu}(1|1)^2_{\ce}$ ($I=3$), and hence the full $\alg{psu}(1|1)^4_{\ce}$. The reference state for the $\alg{psu}(1|1)^2_{\ce}$ ABA is 
\begin{equation}
    \ket{\phi_{p_1}\dots\phi_{p_{N_0}}}\,.
\end{equation}
The monodromy matrix $\aba{M}^1$ is defined as
\begin{equation}
    \aba{M}^1(p_0|\vec{p})=R^{\sL\sL}_{0N}(x_{p_0}^{\pm},x_{p_N}^{\pm})\dots R^{\sL\sL}_{01}(x_{p_0}^{\pm},x_{p_1}^{\pm}),
\end{equation}
with   $R^{\sL\sL}$ given in equation~\eqref{eq:RLLmixed-RtLtLmixed}. Above, the auxiliary space carries momentum $p_0$ and the sites of the physical sites carry momenta $\vec{p}=\{p_i\}_{i=1}^N$.

The auxiliary Bethe equation arises from requiring that the extra term $\ket{X}$ in the eigenstate condition~\eqref{eq:ABA-eigenstate-X} vanish. In order to write these terms explicitly, we need to use the RTT relation 
\begin{equation}\label{RTTsu112}
    R^{\sL\sL}(x_{p_0}^{\pm},x_{p_{0^{\prime}}}^{\pm})\aba{M}^1(p_0|\vec{p})\aba{M}^1(p_0^{\prime}|\vec{p})=\aba{M}^1(p_0^{\prime}|\vec{p})\aba{M}^1(p_0|\vec{p})R^{\sL\sL}(x_{p_0}^{\pm},x_{p_{0^{\prime}}}^{\pm})\,,
\end{equation}
where $0,0^{\prime}$ label two auxiliary spaces with momenta $p_0,p_{0^{\prime}}$.
The commutation relations between operators ($\aba{A}^1,\aba{B}^1$), and ($\aba{D}^1,\aba{B}^1$) follow from~\eqref{RTTsu112} are
\begin{equation}\label{a-bd-bFCR}
    \begin{split}
        \aba{A}^1(p_0|\vec{p})\aba{B}^1(p_{0^{\prime}}|\vec{p})&=\frac{1}{D_{p_{0^{\prime}}p_0}}\aba{B}^1(p_{0^{\prime}}|\vec{p})\aba{A}^1(p_0|\vec{p})-\frac{E_{p_{0^{\prime}}p_0}}{D_{p_{0^{\prime}}p_0}}\aba{B}^1(p_0|\vec{p})\aba{A}^1(p_{0^{\prime}}|\vec{p}),\\
        \aba{D}^1(p_0|\vec{p})\aba{B}^1(p_{0^{\prime}}|\vec{p})&=-\frac{F_{p_0p_{0^{\prime}}}}{D_{p_0p_{0^{\prime}}}}\aba{B}^1(p_{0^{\prime}}|\vec{p})\aba{D}^1(p_0|\vec{p})+\frac{C_{p_0p_{0^{\prime}}}}{D_{p_0p_{0^{\prime}}}}\aba{B}^1(p_0|\vec{p})\aba{D}^1(p_{0^{\prime}}|\vec{p})\,,
    \end{split}
\end{equation}
where the coefficients in the right-hand side are entries of the R-matrix $R^{\sL\sL}$ in equation~\eqref{eq:RLLmixed-RtLtLmixed} . 
Subtracting the second equation from the first one in~\eqref{a-bd-bFCR} and substituting the coefficients from equation~\eqref{ABCDEFcoeffsLrep}, we get
\begin{equation}\label{TL-BL-FCR}
\aba{T}^1(p_0|\vec{p})\aba{B}^1(p_{0^{\prime}}|\vec{p})=\frac{x_{0}^+-x_{0^{\prime}}^-}{x_0^--x_{0^{\prime}}^-}\sqrt{\frac{x_0^-}{x_0^+}}\aba{B}^1(p_{0^{\prime}}|\vec{p})\aba{T}^1(p_0|\vec{p})+\frac{x_{0^{\prime}}^--x_{0^{\prime}}^+}{x_0^--x_{0^{\prime}}^-}\sqrt{\frac{x_0^-}{x_0^+}}\frac{\eta_0}{\eta_{0^{\prime}}}\aba{B}^1(p_{0}|\vec{p})\aba{T}^1(p_{0^{\prime}}|\vec{p}),
\end{equation}
where $x_0^{\pm},x_{0^{\prime}}^{\pm}$ depend on the auxiliary momenta $p_0,p_{0^{\prime}}$ respectively via equation~\eqref{LZhukovskitomommixed}. As discussed in section~\ref{sec:ABA} (see footnote~\ref{ft:massive-aux}) it is simplest to keep the mass parameter of the auxiliary variable non-zero.

The above FCR implies that $\ket{X}$ can be expanded in the basis
\begin{equation}
    \ket{p_0,\hat{y}_{1,i}}\equiv\aba{B}^1(p_0) \aba{B}^{1}(y_{1,1}) \dotsb \aba{B}^{1}(y_{1,i-1})\aba{B}^{1}(y_{1,i+1})\dotsb \aba{B}^{1}(y_{1,{N_1}})  \ket{\phi_{p_1} \dotsb \phi_{p_{N_0}}}
\end{equation}
as
\begin{equation}\label{unwantedbasiscoeffs}
    \ket{X}=\sum_{i=1}^{N_1}M_i\ket{p_0,\hat{y}_{1,i}}\,.
\end{equation}
The coefficients $M_i$ in equation~\eqref{unwantedbasiscoeffs} can be obtained by performing the following set of steps. First, we commute $\aba{B}^{1}(y_{1,i})$ through to the front of the string of $\aba{B}$ operators by using the FCR $\aba{B}^1(p)\aba{B}^1(q)=\frac{F_{pq}}{A_{pq}}\aba{B}^1(q)\aba{B}^1(p)$ which follows from the RTT relation in equation~\eqref{RTTsu112}. The next step involves commuting the transfer matrix $\aba{T}^1(p_0)$ through $\aba{B}^{1}(y_{1,i})$ and picking up the contribution from the second term on the right-hand side of equation~\eqref{TL-BL-FCR} which involves swapping of the auxiliary parameters of $\aba{T}^1$ and $\aba{B}^1$. Next, we commute the $\aba{T}^1$ through the rest of the $\aba{B}^1$ string of operators without any further swapping of the auxiliary parameters, thus picking up contribution from the first term on the right-hand side of equation~\eqref{TL-BL-FCR}.  Finally, we act with the $\aba{T}^1$ operator on the pseudo-vacuum.

Following the above steps, we end up with the coefficient $M_i$ of $\ket{p_0,\hat{y}_{1,i}}$ in the expansion of $\ket{X}$ in equation~\eqref{unwantedbasiscoeffs} as
\begin{equation}\label{Oi-coeffs}
M_i=\left(\prod_{j=1}^{i-1}\frac{F_{q_{1,j}q_{1,i}}}{A_{q_{1,j}q_{1,i}}}\right)\left(\frac{C_{p_0q_{1,i}}}{D_{p_0q_{1,i}}}\right)\prod_{j\neq i}^{N_1}\left(-\frac{F_{q_{1,i}q_{1,j}}}{D_{q_{1,i}q_{1,j}}}\right)\Lambda^1_0(q_{1,i}|\vec{p})\,,
\end{equation}
where $\Lambda^1_0(q|\vec{p})$ is the eigenvalue of the pseudo-vacuum $\ket{\phi_{p_1} \dotsb \phi_{p_{N_0}}}$ under $\aba{T}^1(q|\vec{p})$
\begin{equation}\label{vacuumeigenvalue}
\Lambda^1_0(q|\vec{p})=\prod_{i=1}^{N_0}A_{qp_i}-\prod_{i=1}^{N_0}D_{qp_i}=1-\prod_{i=1}^{N_0}\sqrt{\frac{x^+_{i}}{x^-_{i}}}  \frac{x_q^--x^-_{i}}{x_q^--x^+_{i}}\,.
\end{equation}
The coefficients $M_k$ will \textit{all} be zero simultaneously only if $\Lambda^1_0(q_{1,i}|\vec{p})$ vanishes
\begin{equation}\label{auxconstraintsu112}
   \Lambda^1_0(q_{1,i}|\vec{p})=0,\quad \forall\,\,i=1,2,\dotsc,N_1\,.
\end{equation}
Re-labelling the Zhukovski variables $x_{q_{1,i}}^-=y_{1,i}$ to match the convention in the main text, we recover the auxiliary Bethe equation~\eqref{eq:BE-aux}. Note, the auxiliary Zhukovski variables $x_{q_{1,i}}^+$ can be obtained by using shortening condition in equation~\eqref{eq:momentum-shortening}. However, this is unnecessary for our purpose, as $x_{q_{1,i}}^+$ only show up as overall factors in the Bethe states as discussed in footnote~\ref{fn:B-normalisation}.

\section{\texorpdfstring{$N_0=3$, $N_1=N_3=1$  state from equation~\eqref{N03hws}}{N0=3, N1=N3=1 state}}\label{N03N11N31hws}
Here, we write down explicitly the highest-weight Bethe state with three sites at level $N_1=N_3=1$, in terms of the basis kets
\begin{multline}\label{N11N31hwsbasis}
    \aba{B}^1(y_*) \aba{B}^3(y_*) \ket{\chi_{p_1} \chi_{p_2} \chi_{p_3}}\propto \alpha^2\ket{\tilde{\chi}_{p_1} \chi_{p_2} \chi_{p_3}}+\beta^2 \ket{\chi_{p_1} \tilde{\chi}_{p_2} \chi_{p_3} }
    +\gamma^2\ket{\chi_{p_1} \chi_{p_2} \tilde{\chi}_{p_3}}\\
    -i\alpha\beta\left(\ket{T^1_{p_1} T^2_{p_2} \chi_{p_3}}+\ket{T^2_{p_1} T^1_{p_2} \chi_{p_3}}\right)
    -i\alpha\gamma\left(\ket{T^1_{p_1}\chi_{p_2} T^2_{p_3} }-\ket{ T^2_{p_1}\chi_{p_2} T^1_{p_3}} \right)\\
    -i\beta\gamma\left(\ket{\chi_{p_1}T^1_{p_2} T^2_{p_3} }+\ket{\chi_{p_1} T^2_{p_2} T^1_{p_3}} \right)
    \,,
\end{multline}
where coefficients $\alpha\,,\beta\,,\gamma$ are
\begin{equation}
    \alpha=\frac{\eta_1}{y_{*}-x_1^+}\,,\quad \beta=\frac{\eta_2}{y_{*}-x_2^+} D_{q_{*}p_1}\,,\quad \gamma=\frac{\eta_3}{y_{*}-x_3^+} D_{q_{*}p_1}D_{q_{*}p_2}\,,
\end{equation}
with $ y_{*}=x^-(q_{*})$ defined in equation~\eqref{eq:N0-3-aux-root}, and $\eta_i\,,D_{q_{*}p}$ defined in equation~\eqref{def:etaL},~\eqref{ABCDEFcoeffsLrep} respectively.

\section{Pure RR limit: relativistic ABA}\label{appendix:pureRR}
In this appendix, we will discuss the pure RR limit of our results from the main text in terms of a different spectral parameter $\gamma$. This alternate parameterisation manifests the relativistic invariance present in the pure RR limit, and allows for compact expressions for the Bethe wavefunctions to be written down. We shall review the $\alg{psu}(1|1)^4_{\ce}$ ABA in the relativistic variables and then compute the zero-momentum states that give rise to the protected spectrum. The subtleties involved with the zero-momentum limit, as discussed in the main text, are inherited by this alternate parameterisation.

The pure RR massless limit of the $\alg{psu}(1|1)^2_{\ce}$ R-matrices from equation~\eqref{eq:RLLmixed-RtLtLmixed}, in relativistic variable $\gamma$, is of \textit{difference form}
\begin{equation}\label{RR}
    R^{\sL\sL}_{\alg{psu}(1|1)^2}(\gamma_1-\gamma_2)=\begin{pmatrix}
    1&0&0&0\\
    0&-b&a&0\\
    0&a&b&0\\
    0&0&0&-1
    \end{pmatrix}
,\end{equation}
\begin{equation}\label{R-tR}
    R^{\tilde\sL\tilde\sL}_{\alg{psu}(1|1)^2}(\gamma_1-\gamma_2)=\begin{pmatrix}
    1&0&0&0\\
    0&-b&-a&0\\
    0&-a&b&0\\
    0&0&0&-1
    \end{pmatrix}
,\end{equation}
with $a$ and $b$ defined in terms of the difference of rapidities $\gamma_{12}=\gamma_1-\gamma_2$
\begin{equation}\label{ab-def}
\begin{split}
    a(\gamma_1,\gamma_2)=a(\gamma_1-\gamma_2)&=\sech\frac{\gamma_1-\gamma_2}{2}\equiv \sech\frac{\gamma_{12}}{2},\\
    b(\gamma_1,\gamma_2)=b(\gamma_1-\gamma_2)&=\tanh{\frac{\gamma_1-\gamma_2}{2}}\equiv \tanh{\frac{\gamma_{12}}{2}}\,.
\end{split}
\end{equation}
The rapidity $\gamma$ is related with momentum $p$ as
\begin{equation}
\gamma=\log\tan\frac{p}{4}\,.
\end{equation}
Note the R-matrices in equation~\eqref{RR},~\eqref{R-tR} match those in equation~\eqref{eq:RLLmixed-RtLtLmixed} only when both momenta $p_1(\gamma_1)\,,\,p_2(\gamma_2)$ lie in the range between 0 and $\pi$. Although it is not going to be sufficient to build the monodromy matrices for level-matched momenta (since we are only considering positive momenta for the sites), it is a very similar setup. Later in the section, we will modify this setup while considering level-matched Bethe states.

Following equation~\eqref{Rmatrixsu114}, the $\alg{psu}(1|1)^4_{\ce}$ R-matrix is obtained by tensoring the above two matrices $R^{\sL\sL}\,,R^{\tilde\sL\tilde\sL}$. If we ignore the level-matching condition for now, we can use the the above two R-matrices to build monodromy matrices, and then perform the same steps for ABA construction as in section~\ref{sec:ABA} to get the transfer matrix eigenstates. The Bethe states generated using the operators $\aba{B}^1$ and $\aba{B}^3$ (see equation~\eqref{ABAsu114}), are now labelled by the Bethe roots in relativistic variables 
\begin{equation}\label{ABAsu114RR}
    \ket{\vec{\gamma};\vec{\beta_1};\vec{\beta}_3}\equiv\prod_{i=1}^{N_3}\aba{B}^3(\beta_{3,i})\prod_{j=1}^{N_1}\aba{B}^1(\beta_{1,j})\ket{\chi_{\gamma_1}..\chi_{\gamma_{N_0}}},
\end{equation}
where $\vec{\gamma}=\{\gamma_i\}$, $\vec{\beta}_I=\{\beta_{I,j}\}$ are the momentum-carrying, and auxiliary rapidities. These auxiliary roots are constrained to satisfy the Bethe equations
\begin{equation}\label{Bethe-eq-su114-RR}
\begin{split}
    e^{-iLp_k}&=(-1)^{N_0-1}\prod_{i\neq k}^{N_0}S^2(\gamma_{kj})\prod_{j=1}^{N_1}\coth\tfrac{\beta_{1,jk}}{2}\prod_{l=1}^{N_3}\coth\tfrac{\beta_{3,lk}}{2}\,,\\ 
    1&=\prod_{i=1}^{N_0}\tanh\tfrac{\beta_{1,ki}}{2},\qquad \qquad\qquad k=1,\dotsc,N_1\,,\\
    1&=\prod_{i=1}^{N_0}\tanh\tfrac{\beta_{3,ki}}{2},\qquad \qquad\qquad k=1,\dotsc,N_3\,.
\end{split}
\end{equation}
Above, we have used the shorthand 
\begin{equation}\label{betashorthandsu114}
    \beta_{I,jk}\equiv\beta_{I,j}-\gamma_k,\qquad\qquad\qquad I=1,3\,,
\end{equation}
and $S(\gamma)$ is the famous Zamolodchikov sine-Gordon scalar factor~\cite{Zamolodchikov:1978xm}, as shown in \cite{Bombardelli:2018jkj}. The proof follows from appendix~\ref{app:Betheeqproofs}, with a change of variables. 

In the rest of the section, we compute some of the Bethe states for a single $\alg{psu}(1|1)^2_{\ce}$(with  $R^{\sL\sL}$ as the R-matrix), both at generic and zero momentum. The protected states are obtained by tensoring these Bethe states with ones coming from the other copy of $\alg{psu}(1|1)^2_{\ce}$ ABA (with  $R^{\tilde\sL\tilde\sL}$ as the R-matrix). To avoid repetition, we will omit this final step since it works out the same way as in the Zhukovski variables (see section~\ref{T4section}).

\subsection{Level-matched \texorpdfstring{$\alg{psu}(1|1)^2_{\ce}$}{psu(1|1)2} ABA}
Here, we modify the above setup to allow negative momenta for the sites. This will allow us to impose level-matching for the low magnon solutions with $N_0=1,2,3,4,$ relevant for our protected states discussion. For two particles with mixed kinematics, i.e. $0<p_1<\pi\,,-\pi<p_2<0$, the R-matrix $R^{\sL\ell}_{\alg{psu}(1|1)^2}(\gamma_1-\gamma_2)$ coming from equation~\eqref{eq:RLLmixed-RtLtLmixed}, again in \textit{difference form}, is a slight modification of $R^{\sL\sL}_{\alg{psu}(1|1)^2}(\gamma_1-\gamma_2)$

\begin{equation}\label{R-mixed}
    R^{\sL\ell}_{\alg{psu}(1|1)^2}(\gamma_1-\gamma_2)=\begin{pmatrix}
    1&0&0&0\\
    0&-b&-ia&0\\
    0&-ia&-b&0\\
    0&0&0&1
    \end{pmatrix},
\end{equation}
where $a,b$ are defined using equation~\eqref{ab-def}, and the rapidity $\gamma_2$, for negative values of momentum $-\pi<p_2<0$, is defined as
\begin{equation} 
\gamma_2\equiv\log \tan \frac{p_2+2\pi}{4}\,.
\end{equation}
$R^{\sL\ell}$ (where $\ell$ denotes $L$ with world-sheet $right$ kinematics) shows up whenever we consider a site with negative momentum $p_i<0$. The monodromy matrix is a product of R-matrices, each of which is either $R^{\sL\sL}$ or $R^{\sL\ell}$ depending on the momenta at the sites
\begin{equation}\label{monodromyR-R}
    \aba{M}(\gamma_0|\vec{\gamma})=R_{0N_0}^{\sL q_{N_0}}(\gamma_0-\gamma_{N_0})R_{0N_0-1}^{\sL q_{N_0-1}}(\gamma_0-\gamma_{N_0-1})\dotsc R_{01}^{\sL q_1}(\gamma_0-\gamma_1)\,,
\end{equation}
where $q_i=\sL \,,\ell$ for momentum $p_i(\gamma_i)>0$ or $<0$.

The $\aba{B}$ operator generating the transfer matrix eigenstates on top of the pseudovacuum is $\aba{B}^1(\gamma_0)$ from equation~\eqref{monodromysu112comp}. The auxiliary Bethe equation satisfied by the auxiliary roots $\gamma_0=\beta$, for system with $k$ particles in mixed kinematics, is slightly modified with respect to equation~\eqref{Bethe-eq-su114-RR} (which holds for all sites with positive momenta) by an overall factor of $(-1)^k$ 
\begin{equation}\label{auxBEpureRRgen}
\prod_{i=1}^{N_0} b(\beta-\gamma_i)=(-1)^k\,.
\end{equation}
For $N_0\geq 2$, this equation has two common solutions at $\beta=-\frac{i\pi}{2},\frac{i\pi}{2}$ which translate to $y=0,\infty$ respectively in Zhukovski variable. Expanding the $\aba{B}$ operator near these values of rapidity $\beta$, it behaves as supercharges of the $\alg{psu}(1|1)^2_{\ce}$ algebra (similar to equation~\eqref{eq:B-asympt-sucharge}). Thus, the highest weight Bethe states of the $\alg{psu}(1|1)^2_{\ce}$ algebra are the ones without any roots at $\beta=\pm \frac{i\pi}{2}$. The momentum-carrying Bethe equation does not see any modification in its form w.r.t equation~\eqref{Bethe-eq-su114-RR}, upto the dressing phase which may get modified. It will still be of difference form in the appropriate variables since the R-matrix is. We leave its further investigation for future works, as it is irrelevant for our discussion here\footnote{It will be a reduction of the known phase from \cite{Borsato:2013hoa}.}.
 
\subsubsection{\texorpdfstring{$N_0=1$}{N0 = 1}}
For a single site with rapidity $\gamma_1$, the auxiliary root $\gamma_0=\beta$ satisfying the level-1 Bethe equation~\eqref{auxBEpureRRgen} gives
\begin{equation}
b(\beta-\gamma_1)=1
\end{equation}
which is solved for
\begin{equation}
\beta=\infty.
\end{equation}
Thus the two Bethe states at generic rapidity $\gamma_1$ are
\begin{equation}
\ket{\phi_{\gamma_1}}, \quad \aba{B}^1(\infty)\ket{\phi_{\gamma_1}}\propto\ket{\psi_{\gamma_1}}.
\end{equation}
Note, the above states are not level-matched for generic $\gamma_1$. The zero-momentum limit
\begin{equation}
p\rightarrow 0^+ \quad\Rightarrow\quad \gamma_1\rightarrow -\infty
\end{equation}
is special, since the auxiliary Bethe equation is trivially satisfied for all values of $\beta$. Following the discussion from section~\ref{T4section}, we pick the auxiliary Bethe root $\beta\rightarrow -\infty$ in this limit. Thus, the zero-momentum 1-magnon Bethe states are
\begin{equation}
\ket{\phi_{-\infty}}, \quad \aba{B}^1(-\infty)\ket{\phi_{-\infty}}\propto\ket{\psi_{-\infty}}.
\end{equation}
Taking tensor product of the above states with ones from the other $\mathrm{psu}(1|1)^2_{\ce}$ we recover states in equation~\eqref{state100}\,,\,\eqref{state111}.

\subsubsection{\texorpdfstring{$N_0=2$}{N0 = 2}}
States with two sites satisfy the level-matching condition $0<p_1=-p_2<\pi\,$. Using equation~\eqref{monodromyR-R}, the monodromy matrix is
\begin{equation}
\aba{M}=R^{\sL\ell}(\gamma_0-\gamma_2)R^{\sL\sL}(\gamma_0-\gamma_1),\qquad \gamma_2=-\gamma_1=\gamma.
\end{equation}
The auxiliary Bethe equation for the rapidity $\gamma_0=\beta$ is obtained by substituting $k=1\,,N=2$ in equation~\eqref{auxBEpureRRgen} 
\begin{equation}\label{N=2auxBErel}
\prod_{i=1}^2 b(\beta-\gamma_i)=-1\,,
\end{equation}
with solutions at
\begin{equation}\label{N=2auxroots}
\beta=\pm \frac{i \pi}{2}\,.
\end{equation}
The corresponding  Bethe eigenstates are
\begin{equation}
\begin{aligned}
&\ket{\phi_{\gamma}\phi_{-\gamma}}\,,
\\
 \aba{B}^1(\tfrac{i\pi}{2})\ket{\phi_{\gamma}\phi_{-\gamma}}\propto&\ket{\phi_{\gamma} \psi_{\gamma}}+\ket{\psi_{\gamma}\phi_{\gamma}},\\
 \qquad       \aba{B}^1(-\tfrac{i\pi}{2})\ket{\phi_{\gamma}\phi_{-\gamma}}\propto&\ket{\phi_{\gamma} \psi_{\gamma}}-\ket{\psi_{\gamma}\phi_{\gamma}}\,,
\\
   \aba{B}^1(\tfrac{i\pi}{2})\aba{B}^1(-\tfrac{i\pi}{2})\ket{\phi_{\gamma}\phi_{-\gamma}}\propto& \ket{\psi_{\gamma}\psi_{-\gamma}}\,.
\end{aligned}
\end{equation}
These states organise into a long $(2|2)$-dimensional multiplet of the $\alg{psu}(1|1)^2_{\ce}$ algebra.

Next, we look at the zero-momentum limit. Following the discussion in section~\ref{T4section}, the auxiliary roots are located at rapidities $\beta=\pm \infty$. The Bethe states at these roots are
\begin{equation}\label{N1=1ABAt}
    \begin{aligned}
\lim_{\gamma\rightarrow -\infty}        \ket{\phi_{\gamma}\phi_{-\gamma}}\propto&\ket{\phi_{-\infty}\phi_{\infty}}\,,\\
\lim_{\gamma\rightarrow -\infty}        \aba{B}^1(\infty)\ket{\phi_{\gamma}\phi_{-\gamma}}\propto&\ket{\psi_{-\infty}\phi_{\infty}}\,,\\
    \lim_{\gamma\rightarrow -\infty}    \aba{B}^1(-\infty)\ket{\phi_{\gamma}\phi_{-\gamma}}\propto&\ket{\phi_{-\infty} \psi_{\infty}}\,,\\
\lim_{\gamma\rightarrow -\infty}        \aba{B}^1(-\infty)\aba{B}^1(\infty)\ket{\phi_{\gamma}\phi_{-\gamma}}\propto&\ket{\psi_{-\infty} \psi_{\infty}}\,,
    \end{aligned}
\end{equation}
Since the auxiliaries generating the Bethe states are at $\pm \infty$ (and not at $\pm i\frac{\pi}{2}$), these states are no longer generated by the action of the supercharges and each of the four states are highest weight states of the SUSY algebra. 

\subsubsection{\texorpdfstring{$N_0=3$}{N0 = 3}}
The level-matched rapidities $\gamma_i(p_i)$ for three sites satisfy can be chosen as
\begin{equation}
  \gamma_1=\log{\tan{\frac{p_1}{4}}},\quad\gamma_2=\log{\tan{\frac{p_2}{4}}},\quad \gamma_3=\log{\tan{\frac{p_3+2\pi}{4}}},
\end{equation}
with
\begin{equation}
  \gamma_3=\log{\frac{\sinh{\frac{-\gamma_1-\gamma_2}{2}}}{\cosh{\frac{\gamma_1-\gamma_2}{2}}}}\,.
\end{equation}
In other words by selecting $p_1>p_2>0$ and $p_3<0$. With this choice, the monodromy matrix in relativistic variables is a product of two $R^{\sL\sL}$ and one $R^{\sL\ell}$ 
\begin{equation}
\aba{M}(\gamma_0|\vec{\gamma})=R^{\sL\ell}(\gamma_0-\gamma_3)R^{\sL\sL}(\gamma_0-\gamma_2)R^{\sL\sL}(\gamma_0-\gamma_1)\,.
\end{equation}
The auxiliary Bethe root $\beta$ satisfies the Bethe equation
\begin{equation}\label{N=3M=1rootsRR}
\prod_{i=1}^3 b(\beta-\gamma_i)=-1\,,
\end{equation}
which has solutions
\begin{equation}
\beta=-\infty\,,\pm \frac{i\pi}{2}\,.
\end{equation}
Dropping the $\gamma_i$ subscripts for brevity, and introducing shorthand 
\begin{equation}
    \gamma_i^{\pm}\equiv\gamma_i\pm \frac{i\pi}{2}
\end{equation}
the corresponding $N_1=1$ Bethe states are
\begin{equation}
\begin{aligned}
 \aba{B}^1(-\infty)\ket{\phi\phi\phi}&\propto ie^{-\frac{\gamma_3}{2}}\ket{\phi\phi\psi}+e^{-\frac{\gamma_2}{2}}\ket{\phi\psi\phi}-e^{-\frac{\gamma_1}{2}}\ket{\psi\phi\phi}\,,
\\
\aba{B}^1(\tfrac{i\pi}{2})\ket{\phi\phi\phi}&\propto 
i\sech(\tfrac{\gamma_1^-}{2})
\ket{\phi\phi\psi}
-
\sech(\tfrac{\gamma_2^-}{2})
e^{-\frac{ip_2}{2}}\ket{\phi\psi\phi}
-
\sech(\tfrac{\gamma_3^-}{2})
e^{-\frac{ip_3}{2}}\ket{\psi\phi\phi}\,,
\\
\aba{B}^1(-\tfrac{i\pi}{2})\ket{\phi\phi\phi}&\propto 
i\sech(\tfrac{\gamma_1^+}{2})
\ket{\phi\phi\psi}
-\sech(\tfrac{\gamma_2^+}{2})
e^{\frac{ip_2}{2}}\ket{\phi\psi\phi}
- \sech(\tfrac{\gamma_3^+}{2})
e^{\frac{ip_3}{2}}\ket{\psi\phi\phi}\,.
\end{aligned}
\end{equation}
The  $N_1=2$ Bethe states are
\begin{equation}
\begin{split}
    &\aba{B}^1(\tfrac{i\pi}{2}) \aba{B}^1(-\infty)\ket{\phi\phi\phi}\propto \sech(\tfrac{\gamma_1^-}{2})\ket{\phi\psi\psi}+ \sech(\tfrac{\gamma_2^-}{2})e^{\frac{ip_1}{2}}\ket{\psi\phi\psi}+ i\sech(\tfrac{\gamma_3^-}{2})e^{-\frac{ip_3}{2}}\ket{\psi\psi\phi} \,,\\
    &\aba{B}^1(\tfrac{-i\pi}{2}) \aba{B}^1(-\infty)\ket{\phi\phi\phi} \propto \sech(\tfrac{\gamma_1^+}{2})\ket{\phi\psi\psi}+ \sech(\tfrac{\gamma_2^+}{2})e^{\frac{-ip_1}{2}}\ket{\psi\phi\psi}+ i\sech(\tfrac{\gamma_3^+}{2})e^{\frac{ip_3}{2}}\ket{\psi\psi\phi} \,,\\
    &\aba{B}^1(\tfrac{i\pi}{2})\aba{B}^1(-\tfrac{i\pi}{2})\ket{\phi\phi\phi}\propto e^{-\frac{\gamma_1}{2}}\ket{\phi\psi\psi}+ e^{-\frac{\gamma_2}{2}}\ket{\psi\phi\psi}+ ie^{-\frac{\gamma_3}{2}} \ket{\psi\psi\phi}  \,,
\end{split}
\end{equation}
while the $N_1=3$ state is
\begin{equation}
\aba{B}^1(\tfrac{i\pi}{2})\aba{B}^1(-\tfrac{i\pi}{2}) \aba{B}^1(-\infty)\ket{\phi\phi\phi}\propto \ket{\psi\psi\psi}\,.
\end{equation}

Care needs to be taken in the normalisation of  states generated by $\aba{B}^1(\frac{i\pi}{2})\aba{B}^1(-\frac{i\pi}{2})$, since naively, this product is zero for generic values of rapidities $\gamma_i$. Such zeros did not arise in the main text, because we kept the auxiliary variable in a massive representation and we dropped $y^+$-dependent normalisation factors (see footnote~\ref{fn:B-normalisation} for details). Normalising the states carefully by removing overall factors of the type mentioned in footnote~\ref{fn:B-normalisation} gives the correct non-vanishing eigenstates of the transfer matrix.

In the zero-momentum limit of the above $N_0=3$ Bethe states the magnon rapidities approach $\gamma_1\rightarrow -\infty$, and $\gamma_2=-\gamma_3\rightarrow -\infty$, while the auxiliary  roots end up at $\beta=-\infty$, $-\infty\,,+\infty$.

\subsubsection{\texorpdfstring{$N_0=4$}{N0 = 4}}
For $N_0=4$ sites with two positive and two negative momenta (labelled $a,i$ respectively), the level-matched rapidities $\gamma_i(p_i)$ satisfy 
\begin{equation}
\gamma_a=\log{\tan{\frac{p_a}{4}}},\quad  \gamma_i=\log{\tan{\frac{p_i+2\pi}{4}}},\quad \text{s.t.}\quad \gamma_4=\log\frac{-e^{\gamma_1}-e^{\gamma_2}-e^{\gamma_3}+e^{\gamma_1+\gamma_2+\gamma_3}}{1-e^{\gamma_1+\gamma_2}-e^{\gamma_2+\gamma_3}-e^{\gamma_1+\gamma_3}},
\end{equation}
with $a=1,2,$ and $i=3,4$. The monodromy matrix is 
\begin{equation}
\aba{M}(\gamma_0|\vec{\gamma})=R^{\sL\ell}(\gamma_0-\gamma_4)R^{\sL\ell}(\gamma_0-\gamma_3)R^{\sL\sL}(\gamma_0-\gamma_2)R^{\sL\sL}(\gamma_0-\gamma_1)\,.
\end{equation}
The auxiliary Bethe equation and its solutions for level-matched rapidities are
\begin{equation}\label{N=4auxBEsolv2}
\prod_{i=1}^4b(\beta-\gamma_i)=1 \quad \Rightarrow \quad \beta=\pm \frac{i \pi}{2},\pm \infty \,.
\end{equation}
Using these we generate 1, 4, 6, 4, and 1 Bethe states with 0, 1, 2, 3, and 4 auxiliary roots that span the 16 dimensional space of states. 
As in the $N_0=3$ case, there are spurious zeroes one needs to take care of, arising from the product of $\aba{B}$ operators at $\frac{i\pi}{2}$, $-\frac{i\pi}{2}$.

The zero-momentum limit is approached by taking $\gamma_1=-\gamma_3=-\infty$, followed by $\gamma_2=-\gamma_4=-\infty$. The four auxiliary roots acting on the zero modes are at $\beta=-\infty\,,-\infty\,,+\infty\,,+\infty.$

\section{Coordinate Bethe ansatz}
\label{app:CBA}

Here we will briefly discuss how the coordinate Bethe ansatz can be used to obtain the wave functions we found using the algebraic Bethe ansatz in the main text. The construction we present here is based on the derivation presented in~\cite{Borsato:2016hud,Borsato:2016xns} but formulated in a language similar to the free fermion construction of~\cite{deLeeuw:2020bgo}.

For simplicity we will restrict ourselves to a system with $\algPSU(1|1)^2_{\ce}$ symmetry and consider excitations transforming in the $\rho_{\sL}$ representation, so that we have a boson $\phi$ and a fermion $\psi$. As in the algebraic Bethe ansatz, our starting point is a reference state where all excitations are of highest weight in their representations
\begin{equation}
  \ket{\phi_{p_1} \dotsb \phi_{p_K}} .
\end{equation}
As in equation~\eqref{eq:Psi-as-permutations}, this state can be made into an energy eigenstate by summing over all the permutations of the momenta by repeatedly acting with the S matrix
\begin{equation}
  \ket{\Psi(\sigma_1,\dotsc,\sigma_K)} = \sum_{\tau \in S_K} e^{i(p_{\tau_1} \sigma_1 + \dotsb + p_{\tau_K} \sigma_K)} \Smat_{\tau} \ket{\phi_{p_1} \dotsb \phi_{p_K}} .
\end{equation}
Here $\Smat_{\tau}$ is a component of the full $K$-particle S matrix which permutes the momenta $p_k \to p_{\tau_k}$. Factorised scattering implies that this S matrix can be written as a product of two-particle S matrices, which have the action (see section~\ref{sec:representations})
\begin{equation}
  \Smat \ket{\phi_p \phi_q} = A_{pq} \ket{\phi_q \phi_p} .
\end{equation}
We impose periodicity through the condition
\begin{equation}\label{eq:CBA-periodicity}
  \ket{\Psi(\sigma_1,\dotsc,\sigma_K)} = \ket{\Psi(\sigma_2,\dotsc,\sigma_K,\sigma_1 + L)} ,
\end{equation}
Collecting terms with the same exponential prefactor, we can compare for example the term coming from the trivial permutation on the left-hand side, with the term coming from the cyclic permutation $(K \dotsb 2 1)$ on the right-hand side, as illustrated in figure~\ref{fig:CBA-permutations}. The above equality then leads to the condition
\begin{equation}\label{eq:CBA-BE-rank-1}
  e^{ip_1L} \prod_{j \neq 1} A_{p_1 p_j} = 1.
\end{equation}
\begin{figure}
  \centering

  \begin{tikzpicture}[
    site/.style={circle,draw,outer sep=0,inner sep=0,minimum width=0.4cm,anchor=center},
    label/.style={outer sep=0,inner sep=0}
    ]
    \useasboundingbox (-3,1.3) rectangle (12,-1.4);
    \begin{scope}[yshift=+0.9cm]
      \node[label] at (1,0) {$\scriptstyle \sigma_1\strut$};
      \node[label] at (2,0) {$\scriptstyle \sigma_2\strut$};
      \node[label] at (3,0) {$\scriptstyle \sigma_3\strut$};
      \node[label] at (4,0) {$\scriptstyle \sigma_4\strut$};
      \node[label] at (5,0) {$\scriptstyle \sigma_5\strut$};
      \node[label] at (6,0) {$\scriptstyle \cdots\strut$};
      \node[label] at (7,0) {$\scriptstyle \sigma_K\strut$};
      \node[label] at (8,0) {$\scriptstyle \sigma_1+L\strut$};

    \end{scope}
    \begin{scope}
      \node[site] at (1,0) {$\scriptstyle 1$};
      \node[site] at (2,0) {$\scriptstyle 2$};
      \node[site] at (3,0) {$\scriptstyle 3$};
      \node[site] at (4,0) {$\scriptstyle 4$};
      \node[site] at (5,0) {$\scriptstyle 5$};
      \node at (6,0) {$\scriptstyle \cdots$};
      \node[site] at (7,0) {$\scriptstyle K$};
      \node[anchor=mid] at (-1,0) {Identity};
    \end{scope}
    \begin{scope}[yshift=-0.9cm]
      \node[site] at (2,0) {$\scriptstyle 2$};
      \node[site] at (3,0) {$\scriptstyle 3$};
      \node[site] at (4,0) {$\scriptstyle 4$};
      \node[site] at (5,0) {$\scriptstyle 5$};
      \node at (6,0) {$\scriptstyle \cdots$};
      \node[site] at (7,0) {$\scriptstyle K$};
      \node[site] at (8,0) {$\scriptstyle 1$};
      \node[anchor=mid] at (-1,0) {$(K\dotsb21)$};
    \end{scope}
  \end{tikzpicture}

  \caption{\label{fig:CBA-permutations}The two configurations used to derive equation~\eqref{eq:CBA-BE-rank-1} from equation~\eqref{eq:CBA-periodicity}. The circled numbers represent the excitations $\phi_{p_1}$, $\phi_{p_2}$, \dots, $\phi_{p_K}$. To find equation~\eqref{eq:CBA-BE-rank-1} we pick out the identity permutation, as in the middle line of the figure, from the state on the left-hand side of~\eqref{eq:CBA-periodicity}, and the permutation $(K\dotsb21)$, as in the last line of the figure, from the state on the right-hand side.}
\end{figure}

In order to describe state which also contain $\psi$ excitations we introduce the creation operator $\gen{Q}_y$ which acts by
\begin{equation}\label{eq:Qhat-action}
  \gen{Q}_y \ket{\phi_p} = \frac{\eta_p}{y - x_p^+} \ket{\psi_p} , \qquad
  \gen{Q}_y \ket{\psi_p} = 0,
\end{equation}
with the coproduct
\begin{equation}\label{eq:Qhat-coproduct}
  \Delta(\gen{Q}_y) = \gen{Q}_y \otimes 1 + \sqrt{\frac{x_p^+}{x_p^-}} \frac{y - x_p^-}{y - x_p^+} \otimes \gen{Q}_y .
\end{equation}
The operator $\gen{Q}_y$ interpolates between the supercharges $\gen{Q}_{\sL}$ (for $y=\infty$) and $\bar{\gen{Q}}_{\sR}$ (for $y=0$), as can be seen from equations~\eqref{eq:coprod} and~\eqref{eq:rep-L}, and satisfies the relations\footnote{Here $\Smat_{i,i+1}$ is the two-particle S matrix acting on any two neighbouring excitations.}
\begin{equation}
  \comm{\gen{Q}_y}{\Smat_{i,i+1}} = 0 , \qquad
  \acomm{\gen{Q}_{y_1}}{\gen{Q}_{y_2}} = 0 .
\end{equation}
The first relation above in particular means that
\begin{equation}
  \Smat_{i,i+1} \gen{Q}_y \ket{\phi_{p_1} \dotsb \phi_{p_i} \phi_{p_{i+1}} \dotsb \phi_K} = A_{p_i p_{i+1}} \gen{Q}_y \ket{\phi_{p_1} \dotsb \phi_{p_{i+1}} \phi_{p_i} \dotsb \phi_K} .
\end{equation}
We can now build excited states by acting with some number of $\gen{Q}_y$ on $\ket{\Psi(\sigma_1,\dotsc,\sigma_K)}$. Since $\gen{Q}_y$ commutes with the two-particle S matrix it does not matter if we act with it before or after summing over all permutations.

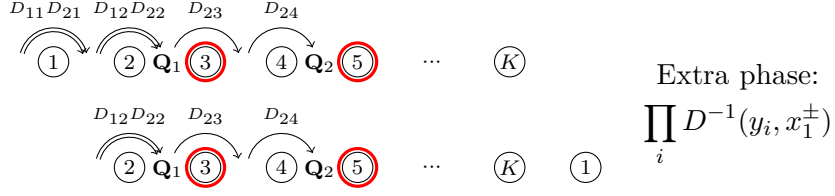
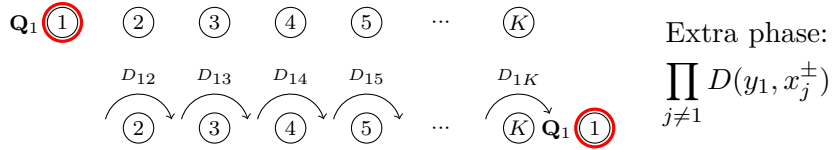
\begin{figure}
  \centering
  \def\centerarc[#1](#2)(#3:#4:#5)
  { \draw[#1] ($(#2)+({#5*cos(#3)},{#5*sin(#3)})$) arc (#3:#4:#5 and #5); }

  \subfloat[\label{fig:CBA-nesting-phases-not-one}Extra phase for creation operators not acting on $\phi_{p_1}$.]{
    \begin{tikzpicture}[
      site/.style={circle,draw,outer sep=0,inner sep=0,minimum width=0.4cm,anchor=center},
      label/.style={outer sep=0,inner sep=0}
      ]
      \useasboundingbox  (-3,1.1) rectangle (12,-2);
      \begin{scope}
        \node[site] at (1,0) {$\scriptstyle 1$};
        \node[site] at (2,0) {$\scriptstyle 2$};
        \node[site] at (3,0) {$\scriptstyle 3$};
        \node[site] at (4,0) {$\scriptstyle 4$};
        \node[site] at (5,0) {$\scriptstyle 5$};
        \node at (6,0) {$\scriptstyle \cdots$};
        \node[site] at (7,0) {$\scriptstyle K$};

        \centerarc[->](1,0)(165:15:0.4)
        \centerarc[->](2,0)(165:25:0.4)
        \draw[red,very thick] (3,0) circle [radius=0.25cm];

        \centerarc[->](1,0)(165:15:0.45)
        \centerarc[->](2,0)(165:25:0.45)
        \centerarc[->](3,0)(155:15:0.45)
        \centerarc[->](4,0)(165:25:0.45)
        \draw[red,very thick] (5,0) circle [radius=0.25cm];

        \node[label] at (0.9,0.5) [anchor=south] {$\scriptscriptstyle D_{11}D_{21}\strut$};
        \node[label] at (2,0.5) [anchor=south] {$\scriptscriptstyle D_{12}D_{22}\strut$};
        \node[label] at (3,0.5) [anchor=south] {$\scriptscriptstyle D_{23}\strut$};
        \node[label] at (4,0.5) [anchor=south] {$\scriptscriptstyle D_{24}\strut$};

        \node[label] at (2.5,-0.0) {$\scriptstyle \gen{Q}_{1}\strut$};
        \node[label] at (4.5,-0.0) {$\scriptstyle \gen{Q}_{2}\strut$};
      \end{scope}
      \begin{scope}[yshift=-1.4cm]
        \node[site] at (2,0) {$\scriptstyle 2$};
        \node[site] at (3,0) {$\scriptstyle 3$};
        \node[site] at (4,0) {$\scriptstyle 4$};
        \node[site] at (5,0) {$\scriptstyle 5$};
        \node at (6,0) {$\scriptstyle \cdots$};
        \node[site] at (7,0) {$\scriptstyle K$};
        \node[site] at (8,0) {$\scriptstyle 1$};

        \centerarc[->](2,0)(165:25:0.4)
        \draw[red,very thick] (3,0) circle [radius=0.25cm];

        \centerarc[->](2,0)(165:25:0.45)
        \centerarc[->](3,0)(155:15:0.45)
        \centerarc[->](4,0)(165:25:0.45)
        \draw[red,very thick] (5,0) circle [radius=0.25cm];

        \node[label] at (2,0.5) [anchor=south] {$\scriptscriptstyle D_{12}D_{22}\strut$};
        \node[label] at (3,0.5) [anchor=south] {$\scriptscriptstyle D_{23}\strut$};
        \node[label] at (4,0.5) [anchor=south] {$\scriptscriptstyle D_{24}\strut$};

        \node[label] at (2.5,-0.0) {$\scriptstyle \gen{Q}_{1}\strut$};
        \node[label] at (4.5,-0.0) {$\scriptstyle \gen{Q}_{2}\strut$};
      \end{scope}
      \node at (10,-0.7) {\parbox{3cm}{\centering Extra phase:\\[4pt]$\displaystyle\prod_i D^{-1}(y_i,x_1^{\pm})$}};
    \end{tikzpicture}
  }

  \subfloat[\label{fig:CBA-nesting-phases-one}Extra phase for creation operators acting on $\phi_{p_1}$.]{
    \begin{tikzpicture}[
      site/.style={circle,draw,outer sep=0,inner sep=0,minimum width=0.4cm,anchor=center},
      label/.style={outer sep=0,inner sep=0}
      ]
      \useasboundingbox  (-3,1.1) rectangle (12,-2);
      \begin{scope}
        \node[site] at (1,0) {$\scriptstyle 1$};
        \node[site] at (2,0) {$\scriptstyle 2$};
        \node[site] at (3,0) {$\scriptstyle 3$};
        \node[site] at (4,0) {$\scriptstyle 4$};
        \node[site] at (5,0) {$\scriptstyle 5$};
        \node at (6,0) {$\scriptstyle \cdots$};
        \node[site] at (7,0) {$\scriptstyle K$};

        \draw[red,very thick] (1,0) circle [radius=0.25cm];

        \node[label] at (0.5,-0.0) {$\scriptstyle \gen{Q}_1\strut$};
      \end{scope}
      \begin{scope}[yshift=-1.4cm]
        \node[site] at (2,0) {$\scriptstyle 2$};
        \node[site] at (3,0) {$\scriptstyle 3$};
        \node[site] at (4,0) {$\scriptstyle 4$};
        \node[site] at (5,0) {$\scriptstyle 5$};
        \node at (6,0) {$\scriptstyle \cdots$};
        \node[site] at (7,0) {$\scriptstyle K$};
        \node[site] at (8,0) {$\scriptstyle 1$};

        \centerarc[->](2,0)(165:15:0.45)
        \centerarc[->](3,0)(165:15:0.45)
        \centerarc[->](4,0)(165:15:0.45)
        \centerarc[->](5,0)(165:15:0.45)
        \centerarc[->](7,0)(165:25:0.45)
        \draw[red,very thick] (8,0) circle [radius=0.25cm];

        \node[label] at (2,0.5) [anchor=south] {$\scriptscriptstyle D_{12}\strut$};
        \node[label] at (3,0.5) [anchor=south] {$\scriptscriptstyle D_{13}\strut$};
        \node[label] at (4,0.5) [anchor=south] {$\scriptscriptstyle D_{14}\strut$};
        \node[label] at (5,0.5) [anchor=south] {$\scriptscriptstyle D_{15}\strut$};
        \node[label] at (7,0.5) [anchor=south] {$\scriptscriptstyle D_{1K}\strut$};

        \node[label] at (7.5,-0.0) {$\scriptstyle \gen{Q}_1\strut$};
      \end{scope}
      \node at (10,-0.7) {\parbox{3cm}{\centering Extra phase:\\[4pt]$\displaystyle\prod_{j\neq1} D(y_1,x_j^{\pm})$}};
    \end{tikzpicture}
  }
  
  \caption{\label{fig:CBA-nesting-phases}Illustration of the extra phases in the nested Bethe equations~\eqref{eq:CBA-BE-nested}. The circled numbers represent the excitations $\phi_{p_1}$, $\phi_{p_2}$, \dots, $\phi_{p_K}$ and the symbols $\gen{Q}_{i}$ represent the creation operator $\gen{Q}_{y_i}$ which acts on the following excitation to the right, as indicated by the thick red circles. Commuting the operator $\gen{Q}_{y_i}$ through the excitation $\phi_{p_j}$ gives a factor $D_{ij} = \sqrt{x_j^+/x_j^-} (y_i - x_j^-)/(y_i - x_j^+)$.}
\end{figure}

Adding excitations to the reference state changes the quantisation condition on the momenta $p_i$. We consider the same types of terms as above, with a trivial permutation on one side and the permutation $(K \dotsb 21)$ on the other side. Let us first look at a term where no $\gen{Q}_y$ operator acts on $\phi_{p_1}$. On the right hand side the $\gen{Q}_y$s do not have to commute through $\phi_{p_1}$ which means we have an extra phase
\begin{equation}
  \prod_{i} \sqrt{\frac{x_{p_1}^-}{x_{p_1}^+}} \frac{y_i - x_{p_1}^+}{y_i - x_{p_1}^-} ,
\end{equation}
as illustrated in figure~\ref{fig:CBA-nesting-phases-not-one}.
Hence we find that periodicity implies the condition
\begin{equation}
  e^{ip_1L} \prod_{j \neq 1} A_{p_1 p_j} \prod_{i} \sqrt{\frac{x_{p_1}^-}{x_{p_1}^+}} \frac{x_{p_1}^+ - y_i}{x_{p_1}^- - y_i} = 1.
\end{equation}
Now consider a term where $\gen{Q}_{y_1}$ acts on $\phi_{p_1}$. The extra factor on the right hand side then takes the form
\begin{equation}
  \prod_{i \neq 1} \sqrt{\frac{x_{p_1}^-}{x_{p_1}^+}} \frac{y_i - x_{p_1}^+}{y_i - x_{p_1}^-} \prod_{j \neq 1} \sqrt{\frac{x_{p_j}^+}{x_{p_j}^-}} \frac{y_1 - x_{p_j}^+}{y_1 - x_{p_j}^-}
  =
  \prod_{i} \sqrt{\frac{x_{p_1}^-}{x_{p_1}^+}} \frac{y_i - x_{p_1}^+}{y_i - x_{p_1}^-} \prod_{j} \sqrt{\frac{x_{p_j}^+}{x_{p_j}^-}} \frac{y_1 - x_{p_j}^+}{y_1 - x_{p_j}^-}
\end{equation}
where the first product has the same origin as in the previous case (for all $\gen{Q}_{y_i}$ with $i \neq 1$), and the second factor comes from commuting $\gen{Q}_{y_1}$ through all $\phi_{p_j}$ with $j \neq 1$, see figure~\ref{fig:CBA-nesting-phases-one}. Compatibility of the resulting equations for $p_1$ then imposes the additional constraint
\begin{equation}
  1 = \prod_{j} \sqrt{\frac{x_{p_j}^+}{x_{p_j}^-}} \frac{y_1 - x_{p_j}^+}{y_1 - x_{p_j}^-} .
\end{equation}

Above we have considered the momentum $p_1$ and the auxiliary parameter $y_1$, but the same conditions of course apply for any other parameters and we find that the Bethe equations take the form
\begin{equation}\label{eq:CBA-BE-nested}
  e^{ip_k L} = \prod_{j \neq 1} A_{p_k p_j}^{-1} \prod_j \sqrt{\frac{x_{p_k}^+}{x_{p_k}^-}} \frac{x_{p_k}^- - y_j}{x_{p_k}^+ - y_j} , \qquad
  1 = \prod_{j} \sqrt{\frac{x_{p_j}^+}{x_{p_j}^-}} \frac{y_k - x_{p_j}^+}{y_k - x_{p_j}^-} .
\end{equation}
These equations exactly match the $\algPSU(1|1)^2_{\ce}$ subsector (\ie, setting $N_3=0$) of the Bethe equations~\eqref{eq:BE-momentum-carrying}, and~\eqref{eq:BE-aux}.

It is now straight forward to obtain wave functions of excited states. For example, the action of a single $\gen{Q}_y$ operator on reference states with two and three excitations are given by
\begin{equation}\label{eq:CBA-N2}
  \gen{Q}_y \ket{\phi_{p_1} \phi_{p_2}}
  =
  \frac{\eta_1}{y - x_1^+} \ket{ \psi_{p_1} \phi_{p_2}}
  + \frac{\eta_2}{y - x_2^+} \frac{y - x_1^-}{y - x_1^+} \sqrt{\frac{x_1^+}{x_1^-}} \ket{ \phi_{p_1} \psi_{p_2}} \,,
\end{equation}
and
\begin{equation}\label{eq:CBA-N3}
  \begin{aligned}
    \gen{Q}_y \ket{\phi_{p_1} \phi_{p_2} \phi_{p_3}} 
    &=
    \frac{\eta_1}{y - x_1^+} \ket{ \psi_{p_1} \phi_{p_2} \phi_{p_3}}
    + \frac{\eta_2}{y - x_2^+} \frac{y - x_1^-}{y - x_1^+} \sqrt{\frac{x_1^+}{x_1^-}} 
    \ket{ \phi_{p_1} \psi_{p_2} \phi_{p_3}}
    \\ &\qquad
    + \frac{\eta_3}{y - x_3^+} \frac{y - x_1^-}{y - x_1^+} \frac{y - x_2^-}{y - x_2^+} \sqrt{\frac{x_1^+}{x_1^-} \frac{x_2^+}{x_2^-}} \ket{ \phi_{p_1} \phi_{p_2} \psi_{p_3} } \,.
  \end{aligned}
\end{equation}
Comparing these expressions with equations~\eqref{eq:n-is-2-single-B-generic} and~\eqref{eq:state-210-general} we see that the states obtained using the $\gen{Q}_y$ operator exactly match those obtained using the $\aba{B}$ operator, up to normalisation.\footnote{The extra signs in equations~\eqref{eq:n-is-2-single-B-generic} and~\eqref{eq:state-210-general} compared to equations~\eqref{eq:CBA-N2} and~\eqref{eq:CBA-N3} appear because the $\algPSU(1|1)^4_{\ce}$ highest weight excitation $\chi$ is a fermion, while we here consider reference states with a bosonic excitation $\phi$.} Similarly, we can check that the two formulation lead to identical wave functions also for states with longer reference states and more creation operators.\footnote{The difference in normalisation of a general state is exactly given by the factor discussed in footnote~\ref{fn:B-normalisation} on page~\pageref{fn:B-normalisation}.}

\bibliographystyle{nb}
\bibliography{refs}

\end{document}